\documentclass[a4,11pt,dvipdfmx]{article}
\usepackage{jheppub}
\usepackage{amsmath,amsfonts,amssymb,bm}
\usepackage{braket}
\usepackage{mathtools}
\usepackage{ulem}

\newcommand{\p}{\partial}
\newcommand{\pslash}{p\kern-1ex /}
\newcommand{\qslash}{q\kern-1ex /}
\newcommand{\lslash}{l\kern-1ex /}
\newcommand{\sslash}{s\kern-1ex /}
\newcommand{\kaslash}{k_a\kern-2ex /}
\newcommand{\kbslash}{k_b\kern-2ex /}
\newcommand{\Dslash}{{\cal D}\kern-1.5ex /}

\newcommand{\beqa}{\begin{eqnarray}}
\newcommand{\eeqa}{\end{eqnarray}}

\newcommand{\bpm}{\begin{pmatrix}}
\newcommand{\epm}{\end{pmatrix}}
\newcommand{\bbm}{\begin{bmatrix}}
\newcommand{\ebm}{\end{bmatrix}}

\newcommand{\bena}{\begin{eqnarray}}
\newcommand{\eena}{\end{eqnarray}}

\def\p{\partial}

\makeatletter

\@addtoreset{equation}{section}
\makeatother
\def\ben{\begin{equation}}
\def\een{\end{equation}}
\def\bena{\begin{eqnarray}}
\def\eena{\end{eqnarray}}

\newcommand{\be}{\begin{align}}
\newcommand{\ee}{\end{align}}
\newcommand{\bea}{\begin{eqnarray}}
\newcommand{\eea}{\end{eqnarray}}
\newcommand{\beas}{\begin{eqnarray*}}
\newcommand{\eeas}{\end{eqnarray*}}

\newcommand{\bdyg}{\mathcal{G}}
\newcommand{\bdyR}{\mathcal{R}}

\newcommand{\pExp}[1]{\langle~#1~\rangle}
\newcommand{\calT}{\mathcal{T}}

\newcommand{\bdyD}{\mathcal{D}}

\def\scriptlap{{\kern1pt\vbox{\hrule height 0.8pt\hbox{\vrule width 0.8pt
  \hskip2pt\vbox{\vskip 4pt}\hskip 2pt\vrule width 0.4pt}\hrule height 0.4pt}
  \kern1pt}}

\def\Biggg#1{{\hbox{$\left#1\vbox to 25pt{}\right.\n@space$}}}
\def\n@space{\nulldelimiterspace=0pt \m@th}
\def\m@th{\mathsurround = 0pt}

\vspace*{0cm}

\title{Instability thresholds for de Sitter and Minkowski spacetimes in holographic semiclassical gravity}

\medskip 

\author[a,b]{Akihiro Ishibashi,}
\author[c]{Kengo Maeda,}
\author[d]{Takashi Okamura}

\affiliation[a]{\it Department of Physics, Nagoya University, Nagoya 464-6802, Japan}

\affiliation[b]{\it Kobayashi-Maskawa Institute, Nagoya University, Nagoya 464-8602, Japan}

\affiliation[c]{\it Faculty of Engineering, Shibaura Institute of Technology, Saitama 330-8570, Japan}

\affiliation[d]{\it Department of Physics and Astronomy, Kwansei Gakuin University, Sanda, Hyogo, 669-1330, Japan}

\emailAdd{ishibashi.akihiro.r7@f.mail.nagoya-u.ac.jp}\emailAdd{maeda302@sic.shibaura-it.ac.jp}\emailAdd{tokamura at kwansei.ac.jp}

\abstract{
We study the stability of $d$-dimensional ($d=3,4,5$) de Sitter and Minkowski spacetimes within the framework of semiclassical gravity sourced by a strongly coupled quantum field with a gravity dual.
Our stability results are derived from a careful analysis of the $d$-dimensional Lichnerowicz equation
with mass-squared $m^2$ and of semiclassical equations involving the dimensionless parameter $\gamma_d$.
For $d=3$, we find that Minkowski spacetime is always unstable against perturbations, whereas de Sitter spacetime becomes stable when a dimensionless parameter $\gamma_3$ exceeds a critical value. In $d=4$, both de Sitter and Minkowski spacetimes become unstable when the parameter $\gamma_4$ exceeds its critical value. In contrast, in $d=5$, de Sitter and Minkowski spacetimes remain stable for almost all values of the parameter $\gamma_5$, except for a regime in which higher-curvature corrections become comparable to the Einstein tensor.
}%


\preprint{NU-QG-13}

\date{}

\begin{document}

\maketitle

\section{Introduction}\label{sec:1}
The stability of Minkowski, de Sitter, and anti-de Sitter (AdS) spacetimes against quantum fluctuations has long been a central issue in quantum gravity, as these spacetimes are maximally symmetric---hence the most fundamental---and have various important applications, e.g., in inflationary cosmology and holographic principle, where quantum effects play essential roles.
A tractable approach to tackling this problem is the so-called semiclassical gravity, in which spacetime is treated classically while matter fields quantum mechanically via the semiclassical Einstein~(SCE) equations.
For example, it was shown that four-dimensional Minkowski spacetime becomes unstable under conformally invariant massless free scalar field when the semiclassical equations contain a fourth-order derivative of the metric with a specific sign~\cite{Horowitz:1978fq}. This analysis was later extended to general massless quantum fields, leading to the conclusion that four-dimensional Minkowski spacetime is generically unstable~\cite{Horowitz:1980fj} (see also~\cite{Suen:1989bg, Suen:1988uf}. For careful treatment of higher curvature terms, see \cite{Simon:1990ic,Simon:1990jn,Simon:1991bm,Parker:1993dk,Anderson:2002fk,Anderson:2009ci}). Despite these studies, several important questions still remain: (i) What happens in other spacetime dimensions or in different curved backgrounds? (ii) Do such instabilities also arise in the presence of strongly interacting quantum matter fields?

Motivated by these considerations, we have previously studied the semiclassical instability of $d$-dimensional AdS spacetime against strongly coupled quantum fields in the framework of holographic semiclassical gravity~\cite{Ishibashi:2023luz,Ishibashi:2023psu,Ishibashi:2024fnm}. By exploiting the holographic methods, we can analyze the above key questions (i) and (ii) in a simple yet interesting example of the AdS spacetime.
More concretely, our strategy is as follows. In the holographic setting, the SCE equations are encoded in mixed boundary conditions at the $d$-dimensional conformal boundary of the $(d+1)$-dimensional AdS bulk spacetime~\cite{Compere:2008us,Ishibashi:2023luz}. Then the perturbed bulk Einstein equations reduce to a set of equations: an equation for a single scalar field toward the bulk radial direction (see (\ref{xi_equation}) below) and a $d$-dimensional Lichnerowicz equation with a mass-squared $m^2$ along the $d$-dimensional conformal boundary spacetime~(see (\ref{H_equation}) below). Our analysis involves a dimensionless parameter $\gamma_d$ which consists of the bulk and boundary Newton couplings, $G_{d+1}$ and $G_d$, and the curvature length, $L$ and $\ell$, respectively (as well as other parameters, e.g., higher curvature coupling constants $\alpha_i$). From the perturbed SCE equations~(see (\ref{semi_Eqs}) below), we can find the algebraic relation between $m^2$ and $\gamma_d$. Then, by examining the Lichnerowicz equation in the allowed range of $m^2$ and $\gamma_d$, we can show the semiclassical (in)stability of the $d$-dimensional boundary spacetime.
For example, we showed in Ref.~\cite{Ishibashi:2023luz} that the three-dimensional AdS (BTZ) solution is unstable under perturbations when the dimensionless constant $\gamma_3$, proportional to the gravitational constant $G_3$, exceeds a critical value $\gamma_{3*}$. This instability arises because the Lichnerowicz equation admits a mode with negative mass-squared $m^2<0$ even when $m^2$ is bounded from below by the Breitenlohner-Freedman (BF) bound~\cite{Breitenlohner:1982jf}. The analysis was further extended to $d=4$ and $d=5$ AdS spacetimes, where a similar instability was found to occur in AdS background with hyperbolic chart in certain range of the parameters $\alpha_i(i=1,2,3)$ characterizing general quadratic theories of gravity~\cite{Ishibashi:2024fnm}
(see also~\cite{Matsui:2018iez, Starobinsky:1980te, Vilenkin:1985md, Chesler:2020exl} for de Sitter case, and \cite{Ghosh:2023gvc} for the semiclassical instabilities of maximally symmetric spacetime in $d=4$ case).

In this paper, we will study the semiclassical (in)stability of $d$-dimensional ($d=3,4,5$) de Sitter and Minkowski spacetimes coupled to a strongly interacting quantum field by applying the above holographic strategy. By doing so, we can exhaust the holographic stability analysis for all maximally symmetric spacetimes. We will find that depending on the values of the mass-squared $m^2$ and the dimensionless constant $\gamma_d$, the de Sitter and Minkowski spacetime can be semiclassically unstable. 
For example, we will show that $3$-dimensional Minkowski spacetime is always unstable when $m^2<0$, whereas $3$-dimensional de Sitter spacetime becomes unstable only when $\gamma_3$ is below some critical value $\gamma_{3*}$. 
For $d=4$ and $d=5$ case, we will find semiclassical instability of de Sitter and Minkowski spacetimes in certain ranges of the parameter values. In Minkowski spacetime, we find that the Lichnerowicz equations always admit an unstable mode whenever the mass-squared takes a negative value. 
For de Sitter case, we find that the details of the stability result depend on the choice of coordinate charts. In the static chart, we show that a large negative mass-squared induces an exponentially growing unstable mode, whereas in the cosmological charts, the perturbations exhibit a power-law behavior that grows indefinitely toward the future whenever the mass-squared is negative, regardless of the type of perturbations (i.e., scalar, vector, or tensor-type). 
In particular, in the global chart, regular initial Cauchy data on a spatially compact slice evolves into such an instability. 
We will briefly discuss this apparent dependency of the stability results on the choice of charts in Subsec.~\ref{subsec:stability:dS}. 

Before going into our stability analysis, we briefly summarize in Table~\ref{table:1} our main results obtained in this paper for de Sitter and Minkowski spacetimes and also for AdS spacetime obtained in our previous papers~\cite{Ishibashi:2023luz,Ishibashi:2023psu,Ishibashi:2024fnm}.  
\begin{table}[h]
\centering
\setlength{\tabcolsep}{0.3pt}
\caption{\small Summary of our stability results of de Sitter/Minkowski/AdS in $d=3,4,5$. The onset of instability is determined by the dimensionless parameter introduced below by (\ref{def_gamma_d}). 
}

{\small 
\label{table:1}
\begin{tabular}{| l|c|c|c|c|c}\hline
  
 {                          } &{} &{}  $d=3$      &{}  $d=4$                         &{}  $d=5$                \\[5pt] \hline \hline
\,\, \,\, deSitter \,\, &{} &{} \,\, stable for $ \gamma_3 \ge \gamma_{3*}$ \,\, 
                                                             &{} \,\, stable for $ \gamma_4 \le \gamma_{4*}$ \,\, 
                                                                                                &{} \,\, stable for $\gamma_5>\gamma_{5*}$ (*) \,\,  \\[5pt]  
{                         } &{}   &{}  unstable for $\gamma_3<\gamma_{3*}$    &{} unstable for 
$\gamma_4>\gamma_{4*}$ 
                                                                                                &{} \,\, unstable 
                                                                                                for $0<\gamma_5\le \gamma_{5*}\ll 1$ 
                                                                                                (*) \, \, \\ [5pt] 
                                                                                                \hline 
\,\,\,\,Minkowski \,\,&{} &{} \,\, always unstable \,\, &{} \,\, stable for $\gamma_4<\gamma_{4*}$ \,\,      &{} \,\, stable for $\gamma_5>\gamma_{5*}$ \,\, \\ [5pt] 
{                         } &{}   &{}    &{}  unstable for $\gamma_4\ge \gamma_{4*}$     &{} unstable for $0<\gamma_5\le \gamma_{5*}\ll 1$ (*) \,\,  \\ [5pt] \hline
   
\,\,\,\,Anti-deSitter \,\, &{} &{} \,\, stable for $ \gamma_3 < \gamma_{3*}$ \,\, 
                                                           &{} \,\, always stable \,\,              &{} \,\, stable for $\gamma_5< \gamma_{5*}$ \,\,   \\[5pt]

{                               } &{}  &{} \,\, unstable for $ \gamma_3\ge \gamma_{3*}$ \,\, 
                                                    &{} \,\, when $|\hat{\alpha}_i|$ is small enough \,\,
                                                                                                  &{} \,\, unstable for $\gamma_5\ge \gamma_{5*}$ (**) \,\,  \\[5pt]\hline  
\end{tabular}
} 
\end{table}

\vspace{-3mm} 
Some comments on Table~\ref{table:1} are in order: 
\begin{itemize}
\item  (*) In $d=5$ both de Sitter and Minkowski spacetimes, unstable modes arise for $0<\gamma_5<\gamma_{5*}\ll 1$, but these modes 
appear only in the regime where the higher-curvature corrections become comparable to the Einstein tensor. 
In this regime, the perturbative treatment breaks down. 

\item (**) In $d=5$ AdS case, the hyperbolic AdS solution is unstable. 

\item For $d=4$ case, Ghosh et al. \cite{Ghosh:2023gvc} also investigated the semiclassical stability of maximally symmetric spacetimes via the holographic method. Although the algebraic equation obtained from their perturbed SCE has the same structure as ours, the difference lies in how the renormalization of the parameters with four-dimensional ambiguities is carried out
\footnote{%
The stability of $4$-dimensional de Sitter spacetime was discussed in terms of the set of three parameters $({\tilde \alpha}, {\tilde \beta}_{\rm eff}, GN^2H^2)$ in \cite{Ghosh:2023gvc}. In the present paper, we introduce $(\hat{\alpha}_{\rm inv}, \hat{\beta}_{\rm inv}, \gamma_4)$ (see (\ref{d=4_dS_algebra_a}), (\ref{d=4_dS_algebra_b}), and (\ref{def_gamma_d}) below for the definitions of these three parameters). 
If we assume the relation between the renormalization points $\mu_{\rm G}=\mu e^{-1/4}$ where $\mu_{\rm G}$ denotes the parameter $\mu$ used in \cite{Ghosh:2023gvc}, while 
the $\mu$ in the right-hand side is introduced in the present paper (see (\ref{bulkct})), then the relations between these parameters are given by 
\bena
 {\tilde \alpha} = \dfrac{1}{12} - \dfrac{2}{\pi \gamma_4}
 (\hat{\alpha}^{\rm (inv)} + 2 \hat{\beta}^{\rm (inv)}) \,, \quad 
 {\tilde \beta}_{\rm eff} = - 2\gamma_{\rm E}
 + \frac{1}{2} - \ln\left(\dfrac{\pi^2}{2}\gamma_4 \right)
 -\dfrac{8}{\pi \gamma_4}{\hat \beta}^{\rm (inv)}\,, \quad
 GN^2H^2 = \dfrac{\pi^2}{2}\gamma_4 \,. 
\nonumber
\eena
}.
In our analysis, we explicitly assume that the higher-curvature corrections are sufficiently small. This assumption simplifies our key formula (the algebraic relation between $m^2$ and $\gamma_4$) for the stability analysis, and clarifies that the dimensionless parameter $\gamma_4$ governs the stability of the background spacetime. In particular, in the de Sitter case, we explicitly solve the Lichnerowicz equations in all cosmological charts as well as in the static chart. 

\item In $d=4$ AdS case, the semiclassical solutions with negative mass-squared do not appear for the parameter range 
$|\alpha_i|\ll \ell^2$~(see Fig.~3 in Ref.~\cite{Ishibashi:2024fnm}).   

\end{itemize} 

This paper is organized as follows. Section~\ref{sec:2} briefly reviews the holographic method used to derive the SCE equations. Section~\ref{sec:3} shows that the background de Sitter and Minkowski spacetimes are indeed solutions of the holographic SCE equations. Section~\ref{sec:4} presents the corresponding algebraic equations and 
investigates the conditions under which semiclassical solutions with negative mass-squared exist.
In Section~\ref{sec:5}, we analytically solve the massive Lichnerowicz equations in $d$-dimensional de Sitter spacetimes. 
We also demonstrate that Minkowski spacetime is unstable whenever the mass-squared is negative by explicitly constructing invariant delta functions. Finally, Section~\ref{sec:6} summarizes our results and discusses their implications.
 
\section{The set up}\label{sec:2}
We first summarize the relevant formulas and setup needed to solve
the SCE equations, which are coupled to a strongly interacting quantum field
via the AdS/CFT duality~
\cite{Maldacena:1997re, Gubser:1998bc, Witten:1998qj},
as presented in the next sections.
We aim to construct the metric $\bdyg_{\mu\nu}$
that satisfies the $d$-dimensional SCE equations \cite{BirrellDavies}
with higher curvature corrections $\mathcal{H}^{(i)}_{\mu\nu}~(i=1,2,3)$
\begin{subequations}
\label{semi_Eqs} 
\begin{align}
  & {\cal E}_{\mu\nu} = 8\, \pi\, G_{d}\, \pExp{{\cal T}_{\mu\nu}}, 
\label{semi_tensor} \\
  & {\cal E}_{\mu\nu}
  := \bdyR_{\mu\nu}-\frac{\bdyR}{2}\bdyg_{\mu\nu}+\Lambda_{d}\,\bdyg_{\mu\nu}
  + \alpha_1 \mathcal{H}^{(1)}_{\mu\nu}
  +\alpha_2 \mathcal{H}^{(2)}_{\mu\nu}+\alpha_3 \mathcal{H}^{(3)}_{\mu\nu}, 
\label{eq:def-calE}
\end{align}
\end{subequations} 
where $\alpha_i$ denote free parameters, $\Lambda_d$ is the (renormalized) cosmological 
constant, and $\pExp{{\cal T}_{\mu\nu}}$ represents the vacuum expectation value of the stress-energy 
tensor of the strongly interacting quantum field with a gravitational dual.
The r.h.s. of Eq.~(\ref{semi_tensor}) is derived
from the quadratic gravity action ($d>3$):
\begin{align}
\label{quadratic_gravity}
{\cal S}=\frac{1}{16\pi G_d}\int d^dx\sqrt{-\bdyg}\left(\bdyR-2\Lambda_d+\alpha_1\bdyR^2
+\alpha_2\bdyR^{\mu\nu}\bdyR_{\mu\nu}+\alpha_3\bdyR^{\mu\nu\rho\sigma}R_{\mu\nu\rho\sigma}\right). 
\end{align} 
By varying the action~(\ref{quadratic_gravity}) with respect to $\bdyg_{\mu\nu}$, 
$\mathcal{H}^{(i)}_{\mu\nu}$ ($i=1,2,3$) are derived as 
\begin{subequations}
\label{def:H}
\begin{align}
\label{def:H1}
& \mathcal{H}^{(1)}_{\mu\nu}=2(\bdyR_{\mu\nu}-\bdyD_\mu\bdyD_\nu)\bdyR-
\bdyg_{\mu\nu}\left(\frac{1}{2}\bdyR^2-2 \bdyD^2 \bdyR   \right),
\\
\label{def:H2}
& \mathcal{H}^{(2)}_{\mu\nu} =2\bdyR_{\mu\rho\nu\sigma}\bdyR^{\rho\sigma}
+\bdyD^2 \bdyR_{\mu\nu}
-\bdyD_\mu\bdyD_\nu\bdyR
-\frac{1}{2}\bdyg_{\mu\nu}(\bdyR_{\sigma\rho}\bdyR^{\sigma\rho}- \bdyD^2\bdyR), 
\\
\label{def:H3}
& \mathcal{H}^{(3)}_{\mu\nu}=2\bdyR_{\mu\rho\sigma\tau}{\bdyR_\nu}^{\rho\sigma\tau}
- \frac{\bdyg_{\mu\nu}}{2}\, \bdyR_{\alpha\beta}^{\rho\sigma}\,
  \bdyR^{\alpha\beta}_{\rho\sigma}
+4{\bdyR}_{\mu\rho\nu\sigma}\bdyR^{\rho\sigma}
-4{\bdyR}_{\mu\rho}{\bdyR^{\rho}}_\nu
-2\bdyD_\mu\bdyD_\nu \bdyR+4 \bdyD^2 \bdyR_{\mu\nu}, 
\end{align}
\end{subequations}
where ${\bdyR^{\alpha}}_{\beta\rho\sigma}$ denotes the Riemann tensor, 
$\bdyR^{\sigma\rho}_{\mu\nu}:={\bdyR^{\sigma\rho}}_{\mu\nu}$, and $\bdyD_\mu$ the covariant 
derivative with respect to ${\bdyg}_{\mu\nu}$.

To evaluate $\pExp{{\cal T}_{\mu\nu}}$
within the framework of the AdS/CFT duality~%
\cite{Maldacena:1997re, Gubser:1998bc, Witten:1998qj},
we consider $d+1$-dimensional bulk AdS spacetime in which ${\bdyg}_{\mu\nu}$ is conformal to 
the AdS boundary metric. 
The bulk metric is given by 
\begin{subequations}
\begin{align}
\label{bulk_metric}
 ds_{d+1}^2&=G_{MN}dX^MdX^N \nonumber \\
&=\Omega^{-2}(z)dz^2+g_{\mu\nu}(z,x)dx^\mu dx^\nu \nonumber \\
&=\Omega^{-2}(z)(dz^2+\tilde{g}_{\mu\nu}(z,x)dx^\mu dx^\nu), 
\end{align}
where we impose 
\begin{align}
\lim_{z\to 0}\tilde{g}_{\mu\nu}(z,x)=\bdyg_{\mu\nu}(x) 
\end{align}
at the AdS conformal boundary. 
Depending on the curvature sign $k$~($k=\pm 1, 0$) of the $d$-dimensional conformal 
boundary spacetime~(see (\ref{Maximally_Sym}) below), the conformal factor takes the form
\begin{align}
\label{Omega}
\Omega(z)
 = \left\{
\begin{array}{lll}
 ({\ell}/{L})\sinh ({z}/{\ell}) & (k=1) \\
 ({\ell}/{L})\sin ({z}/{\ell}) & (k=-1)\\
 {z}/{L} & (k=0)
\end{array}
\right., 
\end{align}
\end{subequations}
where $L$ is the bulk AdS curvature radius and $\ell$ the boundary curvature radius.  
With these choices, the bulk Einstein equations
\begin{align}
\label{Einstein_Eqs}
R_{MN}-\frac{1}{2}G_{MN}R-\frac{d(d-1)}{2L^2}G_{MN}=0
\end{align}
are automatically satisfied, provided that $\tilde{g}_{\mu\nu}(z,,x)$ reduces to the maximally symmetric spacetime 
$\bdyg_{\mu\nu}(x)$, whose Riemann tensor takes the form 
\begin{align} 
\label{Maximally_Sym}
\bdyR_{\mu\nu\alpha\beta}=\frac{k}{\ell^2}
(\bdyg_{\mu\alpha}\bdyg_{\nu\beta}-\bdyg_{\mu\beta}\bdyg_{\nu\alpha}) \,, 
\end{align}
so that $\bdyg_{\mu\nu}(x)$ describes de Sitter (for $k=1$), AdS (for $k=-1$), and Minkowski (for $k=0$) spacetimes.   

The bulk action is decomposed into the Einstein-Hilbert action $S_\text{EH}$, the Gibbons-Hawking surface term 
$S_\text{GH}$, and the counter term $S_\text{ct}$ as 
\begin{subequations}
\begin{align}
\label{bulk_action}
   S_\text{bulk}
  &= S_\text{EH}+S_\text{GH}+S_\text{ct} 
\nonumber \\
  &= \int \frac{d^{d+1}X\sqrt{-G}}{16\pi G_{d+1}}
  \left(R(G)+\frac{d(d-1)}{L^2}  \right)
  +\int_{\Sigma_\epsilon} \frac{d^dx\sqrt{-g}}{8\pi G_{d+1}}K+S_\text{ct}, 
\\
\label{ct_action}
   S_\text{ct}
  &= -\int_{\Sigma_\epsilon} \frac{d^dx\sqrt{-g}}{16\pi G_{d+1}}
  \Bigg( \frac{2(d-1)}{L}+\frac{L}{d-2}R(g)
\nonumber \\ 
  &\hspace*{3.0truecm}
  + \dfrac{c_dL^3}{(d-2)^2}
  \left\{ R_{\mu \nu}(g)R^{\mu \nu}(g)- \dfrac{d}{4(d-1)}R^2(g) \right\}
 \Bigg)\,.
\end{align}
\end{subequations} 
Here $c_d$ is given by 
\begin{eqnarray}
  c_d &=& \left\{
\begin{array}{lll}
 0 & \quad d = 3 \\
 - \ln\left(\mu z  \right) & \quad d=4 \\
 \dfrac{1}{d-4} & \quad d=5 
\label{bulkct} 
\end{array}
\right.,  
\end{eqnarray}
where $\mu$ is an arbitrary mass scale related to renormalization in the field theory. 

According to the AdS/CFT dictionary
, the vacuum expectation 
value of the stress-energy tensor $\pExp{{\cal T}_{\mu\nu}}$ is obtained from the bulk on-shell 
action~(\ref{bulk_action}) by imposing Eqs.~(\ref{Einstein_Eqs})
~\cite{deHaro:2000vlm, Balasubramanian:1999re}.
As shown in~\cite{Ishibashi:2024fnm}, the variation 
of the action~(\ref{bulk_action}) with respect to the conformal boundary metric $\bdyg_{\mu\nu}$ yields
\begin{align}
\label{vev_SE_tensor}
 \pExp{\calT_{\mu\nu}}
 &={-}\dfrac{2}{\sqrt{-\bdyg}}\frac{\delta S_\text{bulk}}{\delta \bdyg^{\mu\nu}}
\nonumber \\
&=\lim_{z\to 0}\frac{1}{8\pi G_{d+1}L}\Biggl{[\dfrac{L^2}{(d-2)\Omega^{d-2}}
\Bigl\{\tilde{K}\tilde{K}_{\mu\nu}-\dfrac{\tilde{g}_{\mu\nu}}{2}(\tilde{K}^{\alpha\beta}
\tilde{K}_{\alpha\beta}+\tilde{K}^2})\Bigr\}
\nonumber \\
&\hspace*{0.5truecm}
-L\tilde{g}_{\nu\rho}\left(\frac{L\Omega}{d-2}\frac{\p}{\p z}+1\right)
\frac{{\tilde{K}_\mu}^\rho-{\delta_\mu}^\rho\tilde{K}}{\Omega^{d-1}}
-\frac{(d-1)\tilde{g}_{\mu\nu}}{2\Omega^d} (1-L\Omega')^2  
  \Biggr]
  +\tau^{(d)}_{\mu\nu} \,,  
\end{align}
where $\tilde{K}_{\mu\nu}$ is the extrinsic curvature associated with $\tilde{g}_{\mu\nu}$, defined as 
\begin{align}
\label{ext_curvature}
\tilde{K}_{\mu\nu}:=-\frac{1}{2}\p_z\tilde{g}_{\mu\nu}
\end{align}
and the tensor $\tau^{(d)}_{\mu\nu}$ is explicitly given in the Appendix \ref{tau}. 

\section{The background solutions}\label{sec:3}
In this section, we show that de Sitter and Minkowski spacetimes can be background semiclassical solutions satisfying Eqs.~(\ref{semi_Eqs}), as in the case of AdS spacetime~\cite{Ishibashi:2024fnm}. 
In the case of Minkowski spacetime, both the vacuum expectation value of the stress-energy tensor 
$\pExp{{\cal T}_{\mu\nu}}$ and ${\cal E}_{\mu\nu}$ in Eqs.~(\ref{semi_Eqs}) vanish by Eqs.~(\ref{vev_SE_tensor}) and (\ref{def:H}), irrespective of the dimension $d$.
Therefore, $d$-dimensional Minkowski spacetime is indeed a semiclassical solution of Eqs.~(\ref{semi_Eqs}).

In the case of de Sitter spacetime, the vacuum expectation value of the stress-energy 
tensor $\pExp{{\cal T}_{\mu\nu}}$ or higher curvature correction term $\mathcal{H}^{(i)}_{\mu\nu}$ does not generically vanish. Only exception is the $d=3$ case in which de Sitter spacetime is the background semiclassical 
solutions of Eqs.~(\ref{semi_Eqs}), as $\pExp{{\cal T}_{\mu\nu}}=0$ and the higher curvature corrections are absent. 

In $d=4$ and $5$ dimensions, we assume that the background solution is the de Sitter spacetime whose 
Riemann tensor is given by Eq.~(\ref{Maximally_Sym}) with $k=1$. 
In $d=4$ case, the vacuum expectation value of the stress-energy 
tensor $\pExp{{\cal T}_{\mu\nu}}$ is induced by the Weyl anomaly as 
\begin{align}
\label{d=4_dS_BG}
\pExp{{\cal T}_{\mu\nu}}=-\frac{3L^3\bdyg_{\mu\nu}}{64\pi G_5\ell^4}
\end{align}
from Eqs.~(\ref{vev_SE_tensor}), while the higher curvature corrections $\mathcal{H}^{(i)}_{\mu\nu}$ vanish. 
Thus, substituting Eq.~(\ref{d=4_dS_BG}) into Eqs.~(\ref{semi_Eqs}), the curvature length $\ell$ is 
determined by the cosmological constant $\Lambda_4$ as 
\begin{align}
\label{Lambda_d=4}
\frac{3}{\ell^2}\left(1-\frac{G_4L^3}{8G_5\ell^2}   \right)=\Lambda_4. 
\end{align}
This implies that the curvature length $\ell$ is smaller than the characteristic length $\sqrt{3/\Lambda_4}$ 
with $k=1$. 

In $d=5$ case, $\pExp{{\cal T}_{\mu\nu}}=0$ due to the absence of the Weyl anomaly, but 
the higher curvature corrections $\mathcal{H}^{(i)}_{\mu\nu}$ appear as 
\begin{align}
\label{H_tensor_5D_BG}
  & \mathcal{H}^{(1)}_{\mu\nu}=-\frac{40}{\ell^4}\bdyg_{\mu\nu},
& & \mathcal{H}^{(2)}_{\mu\nu}=-\frac{8}{\ell^4}\bdyg_{\mu\nu},
& & \mathcal{H}^{(3)}_{\mu\nu}=-\frac{4}{\ell^4}\bdyg_{\mu\nu}. 
\end{align}
Inserting Eqs.~(\ref{H_tensor_5D_BG}) into Eqs.~(\ref{semi_Eqs}), we find that the curvature length 
$\ell$ receives small corrections from the characteristic length $\sqrt{6/\Lambda_5}$
by the higher curvature terms as 
\begin{align}
\label{Lambda_d=5}
  & \frac{6}{\ell^2}+\frac{4}{\ell^2}(10\hat{\alpha}_1+2\hat{\alpha}_2+\hat{\alpha}_3)=\Lambda_5,
& & \hat{\alpha}_i:=\frac{\alpha_i}{\ell^2}. 
\end{align}
In both $d=4$ and $5$ cases, de Sitter spacetimes with the Riemann tensor (\ref{Maximally_Sym}) satisfying 
(\ref{Lambda_d=4}) and (\ref{Lambda_d=5}) are the background semiclassical solutions. 

\section{The perturbed solutions with negative mass-squared}\label{sec:4}
In this section, we derive the Lichnerowicz equation for perturbations on the semiclassical background solutions obtained in Sec.~\ref{sec:3}. We then determine the conditions under which the equation admits a mode of negative mass-squared $m^2<0$ (see below Eqs.~(\ref{xi_equation}) and (\ref{H_equation})). For that purpose, before examining the SCE equations~(\ref{semi_Eqs}), we introduce the following dimensionless constant $\gamma_d$,  
\begin{align}
\label{def_gamma_d}
\gamma_d:=\dfrac{LG_d}{\pi G_{d+1}}
\times 
\left\{
\begin{array}{ll}
\left({L}/{\ell}\right)^{d-2} & (k=\pm 1) \\
1 & (k=0)
\end{array}
\right.,  
\end{align}
which characterizes the solutions to ~(\ref{semi_Eqs}) as will be seen below. 

$\gamma_d$ for $k=\pm 1$ case can be derived from the following holographic 
consideration~\cite{Ishibashi:2023luz}. 
Suppose that the boundary conformal field has $N_{\text{dof}}$
\lq\lq degrees of freedom.\rq\rq\ 
Since the boundary curvature length scale is $\ell$,
we can estimate $\pExp{\calT_{\mu\nu}} \sim N_{\text{dof}}/\ell^{d}$. 
Then, the SCE equations~(\ref{semi_Eqs})
relate $\bdyR \sim 1/\ell^2$
and $G_d\, \pExp{\calT_{\mu \nu}}\sim G_d\, N_{\text{dof}}/\ell^{d}$,
implying  $1/\ell^2 \sim G_d\, N_{\text{dof}}/\ell^{d}$, 
and therefore should involve a dimensionless parameter
$\gamma_d \sim G_d\, N_{\text{dof}}/\ell^{d-2}$.
From the AdS/CFT correspondence, we can also estimate that
$N_{\text{dof}} \sim L^{d-1}/G_{d+1}$ 
and hence obtain the dimensionless parameter
$\gamma_d \sim G_d\, N_{\text{dof}}/\ell^{d-2}$ from the relation  
\begin{align}
  & \frac{ G_d }{ \ell^{d-2} }\, \frac{ L^{d-1} }{ G_{d+1} }
  = \frac{ G_d\, L }{G_{d+1}}\, \left( \frac{L}{\ell} \right)^{d-2}
  = \pi\, \gamma_d
 ~.
\end{align}

Now let us expand the conformal metric $\tilde{g}_{\mu\nu}$ as  
\begin{align}
\label{expansion_conformal}
\tilde{g}_{\mu\nu}(z,\,x)=\overline{\bdyg}_{\mu\nu}(x)+\epsilon h_{\mu\nu}(z,\,x)+O(\epsilon^2), 
\end{align}
where an overbar denotes the background quantity, and 
$\epsilon$ is an infinitesimally small parameter.   
We assume that the metric perturbations $h_{\mu\nu}$ satisfy 
\begin{align}
\label{tt_cond}
{h^\mu}_\mu=\overline{\bdyD}^\nu h_{\mu\nu}=0, 
\end{align}
where the indices are raised and lowered by the background metric $\overline{\bdyg}_{\mu\nu}$, and 
$\overline{\bdyD}_\mu$ is the covariant derivative with respect to $\overline{\bdyg}_{\mu\nu}$. By separating variables
as $h_{\mu\nu}=\xi(z)H_{\mu\nu}(x)$, one obtains two perturbed equations from Eqs.~(\ref{Einstein_Eqs}) as 
\begin{align}
\label{xi_equation}
\xi'' -\frac{(d-1)\Omega'}{\Omega}\xi' +m^2\xi =0, 
\end{align}
and 
\begin{align}
\label{H_equation}
\overline{\bdyD}^2H_{\mu\nu}-\frac{2k}{\ell^2}H_{\mu\nu}=m^2H_{\mu\nu}, 
\end{align} 
where the mass-squared term $m^2$ is introduced as a separation constant and where a dash denotes the derivative with respect to $z$. 
In the following, we assume that the mass-squared $m^2$ is real and $\arg(m)=0$, or $=\pi/2$; namely when $m^2 \geq 0$, 
$m\geq 0$ and when $m^2<0$, $m=i|m|$.

\subsection{de Sitter spacetime}\label{subsec:4.1}
In this subsection, we investigate under what conditions the semiclassical solutions exist in de Sitter 
background. By imposing regularity condition at $z=\infty$ for $\xi$ in (\ref{xi_equation}), the solutions 
of (\ref{xi_equation}) are expressed in terms of the hypergeometric function $F$ as  
\begin{subequations}
\begin{align}
\label{xi_sol_dS}
  & \xi=c(-y)^{\frac{d-1}{2}-p}F\left(p-\frac{d-1}{2},\,p+\frac{1}{2},\,2p+1;\frac{1}{y}   \right),
\\
  & y=\frac{1-\cosh ({z}/{\ell})}{2},
\\
  & p:=\sqrt{\frac{(d-1)^2}{4}-\hat{m}^2}, \qquad \hat{m}^2:=m^2\ell^2,  
\label{def-p}
\end{align}
\end{subequations} 
where $c$ is an arbitrary constant. Note that under the above regularity condition at $y=-\infty$~($z=\infty$), 
when $m^2 \le 0$ we have a non-trivial solution $|\xi|<\infty$, but when $m^2>0$ we have only the trivial solution $\xi=0$. 

For the $d=3$ case, from Eqs.~(\ref{vev_SE_tensor}), (\ref{xi_equation}) and (\ref{xi_sol_dS}), we obtain 
\bena
\label{SE_tensor_d=3dS}
\delta \pExp{{{\mathcal T}_\mu}^\nu}
 &=&\lim_{z\to 0}\frac{\epsilon L^2}{8\pi G_4\ell^3}\left[\frac{\ell \xi'}{2\sinh^2\frac{z}{\ell}}
-\frac{\hat{m}^2\xi}{2\sinh\frac{z}{\ell}}
-\frac{\left(1-\cosh\frac{z}{\ell}\right)^2}{\sinh^3\frac{z}{\ell}}\xi    \right]{H_\mu}^\nu
\nonumber \\
&=&-\frac{\epsilon L^2}{16\pi G_4\ell^3}p\,\hat{m}^2{H_\mu}^\nu, 
\eena
where, in the last equality, we normalized $\xi(0)=1$ by a suitable choice of the integration constant $c$. 
Note that as mentioned before, here $\hat{m}^2<0$. 
Substituting Eqs.~(\ref{SE_tensor_d=3dS}) and (\ref{Pert_Ein_dS}) into the perturbed Eqs.~(\ref{semi_Eqs}), 
we obtain the following algebraic relation between $\hat{m}^2$ and $\gamma_3$: 
\begin{align}
\label{d=3_dS_algebra}
\gamma_3=\frac{1}{\pi\sqrt{1-\hat{m}^2}}.  
\end{align}
This implies that mode solutions with $\hat{m}^2<0$ appear only when 
\begin{align}
\label{def:gamma3_crit}
\gamma_3<\dfrac{1}{\pi} =:\gamma_{3*}.
\end{align}
Note that the massless solution with $\hat{m}^2=0$ also satisfies Eqs.~(\ref{semi_Eqs}), since both sides of 
the perturbed equations are proportional to $\hat{m}^2$.

For the $d=4$ case, the (normalized) solution~(\ref{xi_sol_dS}) is expanded near $z=0$ as 
\begin{align}
\label{dS4_expand}
\xi=1+a_1\left(\frac{z}{\ell}\right)^2+a_2\left(\frac{z}{\ell}\right)^4+b_1\left(\frac{z}{\ell}\right)^4
\ln\left(\frac{z^2}{4\ell^2}\right)+\cdots, 
\end{align}
where the coefficients $a_1$, $a_2$, and $b_1$ are given by 
\begin{align}
\label{coeff_a_i_b_1}
& a_1=\frac{9-4p^2}{16}, \qquad b_1=-\frac{16p^4-40p^2+9}{512}, \nonumber \\
& a_2=\frac{(9-4p^2)\left[25-36p^2-12(1-4p^2)
 \left\{\psi\left(p+\frac{1}{2} \right)+\gamma_E\right\}\right]}{3072}, 
\end{align}
where $\psi(x)$ is the polygamma function and $\gamma_E$ is Euler-Mascheroni constant.  
Inserting Eqs.~(\ref{dS4_expand}) and (\ref{pert_tau4}) into Eq.~(\ref{vev_SE_tensor}), one obtains a finite 
vacuum expectation value of the stress-energy tensor, 
\begin{align}
\label{SE_tensor_dS4}
\delta \pExp{{{\cal T}_\mu}^\nu}=-\frac{\epsilon L^3\hat{m}^2}{64\pi G_5\ell^4}
\left[1+(\hat{m}^2-2)\left\{\psi\left(p+\frac{1}{2} \right)+\gamma_E
-\frac{1}{2}\ln\left(4\ell^2\mu^2 \right)\right\}  \right]{H_\mu}^\nu, 
\end{align}
where the logarithmic divergent term in Eq.(\ref{dS4_expand}) is cancelled by the perturbation of 
$\tau^{(4)}_{\mu\nu}$ in Eq.(\ref{pert_tau4}). 

Although this expression involves an ambiguity associated with the arbitrary renormalization scale $\mu$, 
the physical SCE equations~(\ref{semi_Eqs}) should be independent of $\mu$. The key point is that the free parameters $\alpha_i$~($i=1,2,3$) appearing in ${{\cal E}_\mu}^\nu$~(\ref{semi_tensor}) also depend on $\mu$, namely $\alpha_i=\ell^{2}\hat{\alpha}_i(\mu)$, so that the full SCE equations remain invariant under a change of $\mu$. In other words, the logarithmic term in Eq.~(\ref{SE_tensor_dS4}) can be absorbed by redefinition of the parameters $\alpha_i$, so that the redefined parameters $\hat{\alpha}_i^{\mathrm{(inv)}}$ become invariant under the change of the scale $\mu$. 

For $d \geq 4$, the SCE equations (\ref{semi_Eqs}) include terms quadratic in the curvature. The presence of these higher-order curvature terms broadens the solution space of the SCE equations (\ref{semi_Eqs}). Consequently, this broader solution space includes solutions exhibiting pathological behavior, such as runaway solutions. Several prescriptions have been proposed to eliminate these unphysical solutions from the solution 
space~\cite{Simon:1990ic, Simon:1990jn, Simon:1991bm, Parker:1993dk, Anderson:2002fk, Anderson:2009ci}. 
In this paper, we adopt the procedure that physical solutions must be consistently treatable within perturbation framework even when the effects of higher-order derivatives are considered.   

For $d=4$, it should be noted that $\delta \langle {\cal T}_\mu{}^\nu\rangle$ can involve finite higher derivative terms which stem from the curvature squared. As shown in (\ref{H_equation}), $\hat{m}^2:= m^2\ell^2$ represents the order of the derivatives,  
$\hat{m}^2 \sim \ell^2 \bar{\bdyD}^2$. Then, (\ref{SE_tensor_dS4}) shows that 
$\delta \langle {\cal T}_\mu{}^\nu\rangle$ can involve finite higher-derivative terms of order $\hat{m}^4 \sim \ell^4 \bar{\bdyD}^4$. We then proceed as follows. First we collect the terms proportional to $\hat{m}^2$ and $\hat{m}^4$ 
that can be absorbed into the coefficients $\alpha_i= \ell^2 \hat{\alpha}_i(\mu)$, and then subtract these terms from $\delta \langle {\mathcal T}_\mu{}^\nu\rangle$ so that Eq.~(\ref{SE_tensor_dS4}) contains no nonnegative power
term in $\hat{m}$, up to $O(\hat{m}^2)$ in Eq.~(\ref{SE_tensor_dS4}). 
For that purpose, we expand $\psi$ around $\hat{m}^2=-\infty$ as 
\begin{align}
\label{psi_expand}
\psi\left(p+\frac{1}{2}\right)=\frac{1}{2}\ln(-\hat{m}^2)-\frac{7}{6\hat{m}^2}+O(\hat{m}^{-4}).  
\end{align}
Substituting Eq.~(\ref{psi_expand}) into Eq.~(\ref{SE_tensor_dS4}), we obtain 
\begin{align}
\label{SE_tensor_dS4-expanded}
  & \delta \pExp{ \mathcal{T}_\mu{}^\nu }
  = - \frac{\epsilon L^3\, \Hat{m}^2}{64\, \pi\, G_5\, \ell^4}\,
  \left[\, - \frac{1}{6}
    + (\Hat{m}^2 - 2)\, \left\{ \gamma_E - \ln(2\, \ell\, \mu) \right\}
  \, \right]\, H_\mu{}^\nu
\nonumber \\
  &\hspace*{2.0truecm}
  - \frac{\epsilon L^3\, \Hat{m}^2}{64\, \pi\, G_5\, \ell^4}\,
  \left\{ \frac{ \Hat{m}^2 - 2 }{2}\, \ln(- \Hat{m}^2) + O(1/\Hat{m}^2)
  \right\}\, H_\mu{}^\nu. 
\end{align}

\begin{figure}[htbp]
  \begin{center}
  \includegraphics[width=0.75\linewidth,clip]{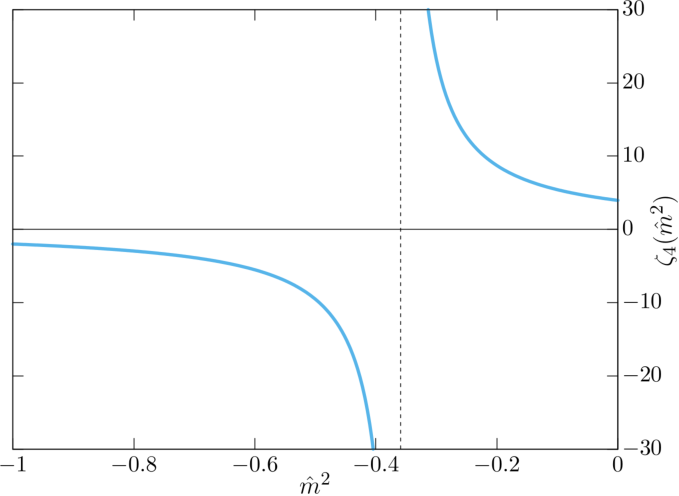}
  \end{center}
  \caption{\small $\zeta_4(\hat{m}^2)$ in Eq.~(\ref{d=4_dS_algebra}) is plotted for 
  $\hat{\alpha}^{\mathrm{(inv)}}=\hat{\beta}^{\mathrm{(inv)}}=0$. 
  The dashed line indicates the position $\hat{m}^2=-0.359$, where the denominator of $\zeta_4(\hat{m}^2)$ vanishes.}
  \label{DS4_algebra}
 \end{figure} 

Combining the nonnegative power terms in the first line of Eq.~(\ref{SE_tensor_dS4-expanded}) with  
the $\hat{\alpha}_i$ terms in the l.~h.~s.~~(\ref{Pert_Ein_dS}) of the perturbed SCE equations, $\delta (\mathcal{E}_\mu{}^\nu)= 8\, \pi\, G_{4}\, \delta \pExp{ \mathcal{T}_\mu{}^\nu }$, 
 \begin{align}
\label{Pert_Ein_dS-main_text}
   \delta (\mathcal{E}_\mu{}^\nu)
  &= - \frac{\epsilon \Hat{m}^2}{2\ell^2}\,
  \left[\, 1
    + 8\, \left\{ 3\, \Hat{\alpha}_1(\mu) + \Hat{\alpha}_2(\mu)
      + \Hat{\alpha}_3(\mu) \right\}
    + (\Hat{m}^2 - 2)\, \left\{ \Hat{\alpha}_2(\mu)
      + 4\, \Hat{\alpha}_3(\mu) \right\}
  \, \right]\, H_\mu{}^\nu
  ~,
\end{align}
we obtain the following algebraic relation among the parameters $\gamma_4$, $\hat{m}^2$, and $\hat{\alpha}_i$: 
\begin{subequations}
\begin{align}
\label{d=4_dS_algebra}
& \gamma_4=\frac{4(1+\hat{\alpha}^{\mathrm{(inv)}}+ \hat{\beta}^{\mathrm{(inv)}}\,\hat{m}^2)}
{\pi\left[\frac{7}{6}+(\hat{m}^2-2)\psi\left(p+\frac{1}{2} \right)\right]} =: \zeta_4(\hat{m}^2), 
\\
& \hat{\alpha}^{\mathrm{(inv)}}:=6(4\hat{\alpha}_1(\mu)+\hat{\alpha}_2(\mu))
+\frac{\pi\gamma_4}{2}\left\{\frac{1}{12}+\gamma_E-\ln(2\ell \mu)\right\},
\label{d=4_dS_algebra_a}
 \\
& \hat{\beta}^{\mathrm{(inv)}}:=\hat{\alpha}_2(\mu)+4\hat{\alpha}_3(\mu)
-\frac{\pi\gamma_4}{4}\left\{ \gamma_E-\ln(2\ell \mu) \right\}, 
\label{d=4_dS_algebra_b}
\end{align}
\end{subequations}
where both the coefficients $\hat{\alpha}^{\mathrm{(inv)}}$ and $\hat{\beta}^{\mathrm{(inv)}}$ should be invariant under the change of $\mu$ and their amplitudes are small enough. 
Since our interest lies in the solutions with negative mass-squared, we have assumed $\hat{m}^2\neq 0$ in deriving Eq.~(\ref{d=4_dS_algebra}). 
It should be noted that the massless case $\hat{m}^2=0$ also satisfies the perturbed equations~(\ref{semi_Eqs}), 
as in the $d=3$ case. 

Let us now examine whether the algebraic equation~(\ref{d=4_dS_algebra}) admits a solution with negative mass-squared $\hat{m}^2<0$. Assuming that the magnitude of the coefficients $|\hat{\alpha}^{\mathrm{(inv)}}|$ and $|\hat{\beta}^{\mathrm{(inv)}}|$ are sufficiently small, Eq.~(\ref{d=4_dS_algebra}) admits a solution in the range $\hat{m}^2\in (-0.359, 0)$, provided that $\gamma_4$ exceeds a critical value $\gamma_{4*}>0$. Here, $\hat{m}^2=-0.359$ denotes the point at which the denominator of $\zeta(x)$ in Eq.~(\ref{d=4_dS_algebra}) vanishes. Figure~\ref{DS4_algebra} shows the function $\zeta(x)$ for $\hat{\alpha}^{\mathrm{(inv)}}=\hat{
\beta}^{\mathrm{(inv)}}=0$. The value of $\gamma_{4*}$ corresponds to the minimum of $\zeta(x)$ in the range $\hat{m}^2\in (-0.359, 0)$, and the qualitative feature of the plot remains unchanged as long as $|\hat{\alpha}^{\mathrm{(inv)}}|$ and $|\hat{\beta}^{\mathrm{(inv)}}|$ are sufficiently small.

For $d= 5$, we substitute (\ref{xi_sol_dS}) and (\ref{pert_tau5}) into (\ref{vev_SE_tensor}), and obtain 
\begin{eqnarray}
\label{SE_tensor_dS5}
\delta \pExp{{{\cal T}_\mu}^\nu}
 &=&-\frac{\epsilon L^4}{144\pi G_6\ell^5}\,
   p\,\hat{m}^2(\hat{m}^2-3){H_\mu}^\nu 
 \nonumber \\
 &=& -\frac{\epsilon L^4}{144\pi G_6\ell^5}(-\hat{m}^2)^{5/2}\left( 1-\dfrac{3}{\hat{m}^2} \right) \sqrt{1-\dfrac{4}{\hat{m}^2}}{H_\mu}^\nu  . 
\end{eqnarray} 
Note that for the $d=5$ case, we cannot apply a similar procedure performed in $d=4$ case---explained just above (\ref{psi_expand}), since terms with odd power $\hat{m}^5$ are involved. 
Combining (\ref{SE_tensor_dS5}) and (\ref{Pert_Ein_dS}), we find that the perturbed SCE equations~(\ref{semi_Eqs}) 
reduce to the following algebraic equation:   
\begin{subequations}
\begin{align}
\label{d=5_dS_algebra}
& \gamma_5=\frac{9(1+ \hat{\alpha}_5+\hat{\beta} \,\hat{m}^2)}{\pi(\hat{m}^2-3)
\sqrt{4-\hat{m}^2}} =:\zeta_5(\hat{m}^2) , \\
& \hat{\alpha}_5 :=8(5\hat{\alpha}_1+\hat{\alpha}_2)-4\hat{\alpha}_3, \qquad  
  \hat{\beta} :=\hat{\alpha}_2+4\hat{\alpha}_3.  
\end{align}
\end{subequations}
Note that the massless solution with $\hat{m}^2=0$ also satisfies perturbed SCE equations~(\ref{semi_Eqs}), since both sides of 
the perturbed equations are proportional to $\hat{m}^2$. In this case, the SCE equations (\ref{semi_Eqs}) 
reduce to the (perturbed) vacuum Einstein equations. 

Since $\hat{\alpha}_5$ is small enough, $1+ \hat{\alpha}_5 >0$. 
This means that Eq.~(\ref{d=5_dS_algebra}) does not allow the solutions with negative mass-squared $\hat{m}^2<0$ for $\hat{\beta} \le 0$. 
For $\hat{\beta}>0$, under the conditions $|\hat{\alpha}_5|\ll 1$ and $|\hat{\beta}|\ll 1$, 
$\zeta$ in Eq.~(\ref{d=5_dS_algebra}) takes a local maximum at 
\begin{align}
\hat{m}^2=\hat{m}_0^2\simeq -\frac{3}{\hat{\beta}}<0, 
\end{align}
with the approximate value, 
\begin{align}
\label{maximum_gamma_5}
\zeta_5(\hat{m}^2_0)\simeq \frac{2\sqrt{3}\, \hat{\beta}^\frac{3}{2}}{\pi}. 
\end{align}
Therefore, the sufficient condition that ensures the absence of a solution with $\hat{m}^2< 0$ in the algebraic Eq.~(\ref{d=5_dS_algebra}) is given by
\begin{align}
\label{cond_gammma_5_dS}
 \hat{\beta} \le 0 \quad \mbox{or} \quad 
\gamma_5\gtrapprox\frac{2\sqrt{3} \hat{\beta} ^\frac{3}{2}}{\pi} =:\gamma_{5*}>0. 
\end{align}
When $\gamma_5$ is sufficiently small, i.e., $\gamma_5<\gamma_{5*}$, a solution with negative 
mass-squared exists. However, such a solution appears only when $|\hat{m}^2|\sim |1/\hat{\beta}|\sim |1/\hat{\alpha}_i|$ 
for some $i$. This indicates that the magnitude of the higher-curvature terms is of the same order as that of the 
Einstein tensor, corresponding to a solution located at the boundary of the unphysical region, where higher-curvature 
terms dominate over the Einstein term in the semiclassical Einstein equations, i.e., 
$|\bdyR_{\mu\nu}-\bdyR \bdyg_{\mu\nu}/2|\ll |\alpha_i \mathcal{H}^{(i)}_{\mu\nu}|$ for at least one $i$. 
Therefore, de Sitter spacetime is stable except for the regime where the perturbative analysis is 
unreliable~\footnote{Even in the case of $d=4$, there exists another critical value $\tilde{\gamma}_{4*}$, 
which is sufficiently small, i.~e., $\tilde{\gamma}_{4*}=O(\hat{\beta}^{\mathrm{(inv)}})$ when 
$\hat{\beta}^{\mathrm{(inv)}}>0$. 
In this situation, an unreliable solution appears for $0<\gamma_4<\tilde{\gamma}_{4*}\ll 1$.}.

\subsection{Minkowski spacetime}
In this subsection, we investigate under what conditions the semiclassical solutions exist in Minkowski 
background. Inserting $\Omega=z/L$ into Eq.~(\ref{xi_equation}), one obtains 
the general solution expressed by the Hankel functions of the first and second kind as  
\begin{align}
\label{Hankel_sol}
\xi(z)=z^\frac{d}{2}\left(c_1H^{(1)}_\frac{d}{2}(mz)+c_2H^{(2)}_\frac{d}{2}(mz)\right) .
\end{align}

Near the Poincar\'{e} horizon located at $z=\infty$ in the bulk, the asymptotic behavior of the Hankel functions 
are  
\begin{subequations}
\begin{align}
\label{Hankel_asym}
& H_\frac{d}{2}^{(1)}(mz)\sim \sqrt{\frac{2}{\pi z}}e^{i\left(mz-\frac{d+1}{4}\pi  \right)}, \\
&  H_\frac{d}{2}^{(2)}(mz)\sim \sqrt{\frac{2}{\pi z}}e^{-i\left(mz-\frac{d+1}{4}\pi  \right)}. 
\end{align}
\end{subequations} 
Having two solutions, we consider boundary conditions: when $m^2>0$, we must take $c_2=0$ because at the horizon 
the first term of (\ref{Hankel_sol}) corresponds to the horizon incoming solution, while when $m^2<0$, taking $\arg(m)=\pi/2$
as mentioned below (\ref{H_equation}), we must again take $c_2=0$ for the regularity condition at $z=\infty$.

When the boundary is Minkowski spacetime, the perturbation of the l.~h.~s.~of the SCE equations~(\ref{semi_Eqs}) is 
obtained from Eqs.~(\ref{formula_dS}) and (\ref{H_pert_dS}) in the limit $\ell\to \infty$. The result is 
\begin{align}
\label{Pert_Ein_Min}
  \delta (\mathcal{E}_\mu{}^\nu)
  = - \frac{ \Tilde{m}^2 }{2L^2}\,
  \left\{ 1 + \left( \Tilde{\alpha}_2
  + 4\, \Tilde{\alpha}_3 \right)\, \Tilde{m}^2 \right\} 
  \, H_\mu{}^\nu,
\end{align}
where 
\begin{align}
\label{def_Min_tilde}
  & \tilde{m}^2:=m^2L^2,
& & \tilde{\alpha}_i:=\frac{\alpha_i}{L^2}. 
\end{align} 

For the $d=3$ case, the stress-energy tensor~(\ref{vev_SE_tensor}) at O($\epsilon$) reduces to 
\begin{align}
\label{d=3_Min_SE}
\delta\pExp{{\cal T}_{\mu\nu}}=\lim_{z\to 0}\frac{\epsilon L^2}{16\pi G_4z}
\left[h''_{\mu\nu}-\frac{h'_{\mu\nu}}{z}\right]=\lim_{z\to 0}\frac{\epsilon L^2}{16\pi G_4z}
\left(-m^2h_{\mu\nu}+\frac{h'_{\mu\nu}}{z}\right) 
\end{align}
by (\ref{xi_equation}). We shall normalize the solution~(\ref{Hankel_sol}) so that $\xi(0)=1$. 
Then, inserting the expansion of $\xi$ at $z=0$
\begin{align}
\label{hankel_series_D3}
\xi=1+\frac{(mz)^2}{2}+\frac{i}{3}(mz)^3+\cdots
\end{align}
into (\ref{d=3_Min_SE}), one obtains 
\begin{align}
\label{SE_tensor_Min3}
\delta \pExp{{{\cal T}_\mu}^\nu}=\frac{i\epsilon L^2}{16\pi G_4}m^3{H_\mu}^\nu. 
\end{align}
It follows from this expression that in order for the expectation value of the stress-energy tensor to be real, the mass-squared must be negative: $m^2<0$.
By Eqs.~(\ref{Pert_Ein_Min}) and (\ref{SE_tensor_Min3}), the perturbed SCE equations~(\ref{semi_Eqs}) reduce to the algebraic equation
\begin{align}
\label{d=3_Min_algebra}
 \tilde{m}^2 =-\left( \frac{G_4}{G_3L}  \right)^2.
\end{align}
Thus, only a single negative mass-squared solution appears. Note that there is another trivial solution with $m=0$, as
both side of SCE equations~(\ref{semi_Eqs}) are proportional to $m^2$.

For the $d=4$ case, the solution~(\ref{Hankel_sol}) is expanded near the AdS boundary $z=0$ as
\begin{align}
\label{expand_D4_Min}
\xi&=\frac{i\pi}{4}\,(mz)^2H_2^{(1)}(mz) \nonumber \\
&=1+\frac{(mz)^2}{4}-\frac{(mz)^4}{64}\left(-3+4\gamma_E-2\pi i+4\ln\left(\frac{mz}{2} \right)+\cdots \right),
\end{align}
where $\gamma_E$ is Euler-Mascheroni constant, and we normalized $\xi$ so that $\xi(0)=1$.

\begin{figure}[htbp]
  \centering
  \includegraphics[width=0.75\linewidth,clip]{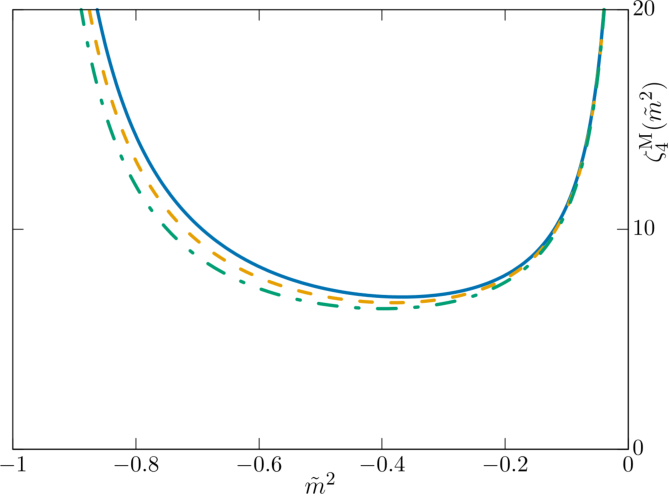}
  \caption{\small $\zeta_4^M(\tilde{m}^2)$ in Eq.~(\ref{d=4_Min_algebra}) is plotted for $\tilde{\beta}^{\mathrm{(inv)}}=0$~(blue, solid),\,
  $0.06$~(orange, dashed), \,$0.12$~(green, dot-dashed). As $\tilde{\beta}^{\mathrm{(inv)}}$ becomes large, the minimum value for $\gamma_4$ to have a solution with $\tilde{m}^2<0$ is lowered. }
  \label{Min4_algebra}
 \end{figure} 

The logarithmic divergence at $z=0$ in Eq.~(\ref{expand_D4_Min}) does not appear in the vacuum 
expectation value of the stress-energy tensor~(\ref{vev_SE_tensor}), as it is cancelled by the logarithmic 
divergence in $\delta\tau_{\mu\nu}^{(4)}$~(\ref{pert_tau4}). Thus, we obtain a finite value of the 
expectation value as 
\begin{eqnarray}
\label{SE_tensor_Min4}
  \delta\pExp{{{\cal T}_\mu}^\nu }
  &=& - \frac{\epsilon L^3\, m^4}{128\, \pi\, G_5}\,
  \left\{ 2\, \gamma_E
  + \ln (-\Tilde{m}^2) - 2\, \ln(2\, \mu\, L) 
  \right\}\, {H_\mu}^\nu
\nonumber \\
  &=&-\frac{\epsilon L^3\, m^4}{128\, \pi\, G_5}\,
  \left\{ 2\, \gamma_E
   - 2\, \ln(2\, \Tilde{\mu}) 
  + \ln |\tilde{m}|^2
  \right\}\, {H_\mu}^\nu \,,
\end{eqnarray}
where we set 
$\tilde{\mu}:=\mu L$. Then, substituting Eqs.~(\ref{Pert_Ein_Min}) and (\ref{SE_tensor_Min4}) into
the perturbed equations~(\ref{semi_Eqs}), and 
following the same procedure as in the $d=4$ de Sitter case, we obtain the algebraic equation 
\begin{subequations}
\begin{align}
\label{d=4_Min_algebra}
  & \gamma_4
  = \frac{8}{\pi}\,
    \left( \Tilde{\beta}^{\mathrm{(inv)}} + \frac{1}{ \Tilde{m}^2 } \right)\,
    \frac{1}{ \ln (- \Tilde{m}^2) }
  =: \zeta^{\text{M}}_4(\tilde{m}^2),
\\ 
  & \tilde{\beta}^{\mathrm{(inv)}}
  := \Tilde{\alpha}_2(\mu) + 4\, \Tilde{\alpha}_3(\mu)
  - \frac{\pi\, \gamma_4}{4}\, \left\{ \gamma_E 
  - \ln(2\, \Tilde{\mu}) 
  \right\},
\end{align} 
\end{subequations} 
where $\tilde{\beta}^{\mathrm{(inv)}}$ is the invariant parameter, which is independent of $\mu$, as in the de Sitter case given in Eq.~(\ref{d=4_dS_algebra}).

Assuming that $\tilde{\beta}^{\mathrm{(inv)}}$ is sufficiently small, a negative mass-squared solution appears in the regime $-1<\tilde{m}^2<0$ when $\gamma_4$ exceeds a critical value $\gamma_{4*}~(>0)$.
Figure~\ref{Min4_algebra} shows the plot of the function $\zeta_4^M$ for various values of
$\tilde{\beta}^{\mathrm{(inv)}}$. Since the denominator of Eq.~(\ref{d=4_Min_algebra}) vanishes at 
$\tilde{m}^2=-1$, the function $\zeta_4^M$ diverges as $\tilde{m}^2\to 0$ and $\tilde{m}^2\to -1$. 
Thus, $\zeta_4^M$ attains a positive minimum value $\gamma_{4*}$ for each small $\tilde{\beta}^{\mathrm{(inv)}}$. This indicates that Eq.~(\ref{d=4_Min_algebra}) admits two negative mass-squared solutions 
when $\gamma_4>\gamma_{4*}$. Note that there is another possibility: the algebraic 
equation~(\ref{d=4_Min_algebra}) may admit a solution with $\tilde{m}^2=-O(1/\tilde{\beta}^{\mathrm{(inv)}})$  
under the condition $\tilde{\beta}^{\mathrm{(inv)}}>0$. However, as in the $d=4$ de Sitter case, 
such a solution is unreliable because the magnitude of the higher-curvature terms is of the same order as 
that of the Einstein tensor.

For the $d=5$ case, the vacuum expectation value of the stress-energy tensor~(\ref{vev_SE_tensor}) 
is calculated by inserting Eq~(\ref{pert_tau5}) and using Eq.~(\ref{xi_equation}) as,   
\begin{align}
\label{SE_tensor_Min5}
\delta\pExp{{\mathcal T}_{\mu\nu}}=\frac{\epsilon L^4}{48\pi G_6}
\left[\frac{3}{z^4}h'_{\mu\nu}-\frac{m^2}{z^3}h_{\mu\nu}-\frac{m^4}{3z}h_{\mu\nu}\right]. 
\end{align}
As before, we shall normalize $\xi$ in (\ref{Hankel_sol}) so that $\xi(0)=1$. So, $\xi$ is expanded as 
\begin{align}
\label{expand_D5_Min}
\xi=\frac{i}{3}\sqrt{\frac{\pi}{2}}(mz)^\frac{5}{2}H^{(1)}_\frac{5}{2}(mz)
=1+\frac{(mz)^2}{6}+\frac{(mz)^4}{24}+\frac{i}{45}(mz)^5+\cdots,  
\end{align}
and (\ref{SE_tensor_Min5}) reduces to 
\begin{align}
\delta \pExp{{{\mathcal T}_\mu}^\nu}=\frac{i\epsilon L^4m^5}{144\pi G_6}{H_\mu}^\nu. 
\end{align}
As mentioned below (\ref {SE_tensor_Min3}), in order for the expectation value of the stress-energy tensor to be real, the mass-squared must be negative: $m^2<0$.
Thus, the SCE equations~(\ref{semi_Eqs}) is transformed into the following algebraic 
equation~\footnote{Note that the SCE equations~(\ref{semi_Eqs}) also admit the massless 
solution, $\tilde{m}=0$ as before.}
\begin{subequations}
\begin{align}
\label{d=5_Min_algebra}
  & \gamma_5
  = \frac{9}{\pi}\, \left( \Tilde{\beta} + \frac{1}{ \Tilde{m}^2 } \right)\,
  \frac{1}{ \sqrt{ - \Tilde{m}^2 } }
  =: \zeta^{\text{M}}_5(\Tilde{m}^2)
  ~, \\
  & \tilde{\beta}:=\tilde{\alpha}_2+4\tilde{\alpha}_3. 
\end{align}
\end{subequations}
When $\tilde{\beta}\le 0$, Eq.~(\ref{d=5_Min_algebra}) clearly possesses
no solution with $\tilde{m}^2<0$. 
For $\tilde{\beta}>0$, $\zeta^{\text{M}}_5(\Tilde{m}^2)$ attains
a maximum at $\Tilde{m}^2 = \Tilde{m}^2_0 := - 3/\Tilde{\beta}$,
with the value
\begin{align}
\label{maximum_gamma_5-Min}
  & \zeta^{\text{M}}_5(\Tilde{m}^2_0)
  = \frac{ 2 \sqrt{3}\, \Tilde{\beta}^{3/2} }{\pi}
  .
\end{align}
So, for a sufficient condition that ensures the absence of a solution with negative mass-squared in the algebraic Eq.~(\ref{d=5_Min_algebra}) is given by
\begin{align}
\label{cond_gammma_5_Min}
\tilde{\beta}\le 0 \qquad \text{or} \qquad \gamma_5>\gamma_{5*}:=
\frac{2\sqrt{3}\tilde{\beta}^\frac{3}{2}}{\pi}.
\end{align} 
As in the de Sitter case, when $\gamma_5$ is sufficiently small, i.e., $\gamma_5<\gamma_{5*}$, a solution with 
negative mass-squared exists. However, such a solution appears only when 
$|\tilde{m}^2|\sim |1/\tilde{\beta}|\sim |1/\tilde{\alpha}_i|$ for some $i$, which corresponds to the regime where the 
higher-curvature terms become comparable to the Einstein tensor. In this regime, the SCE equations 
enter the boundary of the unphysical region, where higher-curvature terms dominate over the 
Einstein term, i.e., $|\bdyR_{\mu\nu}-\bdyR \bdyg_{\mu\nu}/2|\ll |\alpha_i \mathcal{H}^{(i)}_{\mu\nu}|$ 
for at least one $i$. Therefore, Minkowski spacetime is stable except for the regime where the perturbative analysis is 
unreliable. 

\section{Instabilities of de Sitter and Minkowski spacetimes}\label{sec:5}
In the previous section~\ref{sec:4}, we have clarified under what conditions the mode solutions with $m^2<0$ of Eq.~(\ref{xi_equation}) appears. In this section, we demonstrate that such mode solutions $m^2<0$ lead to instability of the background semiclassical solutions. 

\subsection{de Sitter spacetime}\label{subsec:stability:dS}
Let us first investigate the solutions with $m^2<0$ of Eqs.~(\ref{H_equation}) in the de Sitter background spacetime. We first examine perturbations in the static chart, where we state our stability criterion and show the instability by explicitly constructing an unstable mode. Then, we examine perturbations in the cosmological (spatially flat, or closed) charts. 

\subsubsection{Stability analysis in the static region of de Sitter spacetime} 
We consider time-dependent perturbations under the metric ansatz
\bena
\label{dS_metric_Static}
ds_d^2 &=&
-\frac{f(u)}{u}(1+\epsilon T(u)e^{-i\omega t})dt^2
+\frac{\ell}{2u^2 f(u)} \epsilon S(u)e^{-i\omega t} dtdu
\nonumber \\
&{}&
+\frac{\ell^2}{4u^2 f(u)}(1+\epsilon U(u)e^{-i\omega t})du^2
+\frac{\ell^2}{u}(1+\epsilon R(u)e^{-i\omega t})d\Omega^2_{d-2},  
\eena
where $f(u) = u - 1$, and
$d\Omega^2_{d-2}$ is the $d-2$-dimensional unit sphere.
The de Sitter (cosmological) horizon is located at $u=1$. 
Note that after the coordinate transformation $u=\ell^2/r^2$, the metric with $\epsilon=0$ reduces to 
the standard static chart of $d$-dimensional de Sitter spacetime, where $u=\infty$ corresponds to the origin $r=0$. 

Under this metric ansatz, we consider only a restricted class of metric perturbations. 
In general, one can classify generic metric perturbations on the background (\ref{dS_metric_Static}) with $\epsilon=0$ into the tensor, vector, and scalar-type, as given in \cite{Kodama:2000fa}. Based on our previous analysis in the AdS case~\cite{Ishibashi:2024fnm}, which showed that the only scalar-type perturbations can cause instability, in this paper we focus on the scalar-type metric perturbations. Our metric perturbations can be interpreted as a massive tensor field perturbations, as seen in (\ref{H_equation}). It is known that in general, the equations for scalar-type of massive tensor field perturbations do not reduce to a set of decoupled master equations~\cite{Cardoso:2019mes}. For this technical reason, we further need to restrict our scalar-type metric perturbations to the above form of (\ref{dS_metric_Static}), allowing us to derive a single, decoupled master equation, and discuss the stability in a definitive manner
\footnote{ 
When $d \geq 5$, we can consider the tensor-type metric perturbations with respect to the $(d-2)$-sphere, which can be expressed in terms of the transverse-traceless tensor harmonics ${\mathbb T}_{ij}$ defined on $(d-2)$-sphere
and a single scalar $\Phi(t,u)$ as $H_{t \mu}=H_{u \mu}=0$
and 
$H_{ij} = 2\, \left( u/\ell^2 \right)^{(d-6)/4} \Phi(t,u) {\mathbb T}_{ij}$
(which should not be confused with tensor-perturbations
in the cosmological context. See, e.g.,  \cite{Kodama:2000fa}).
For generic modes of the tensor-type perturbations,
our metric perturbation equations (\ref{H_equation}) reduce to the equation
\[
\left( - \partial^2_t + \partial^2_* 
  - \dfrac{f(u)}{u} \left[ \dfrac{u}{\ell^2}
    \left\{ \dfrac{(d-2)(d-4)}{4} + l(l+d-3) \right\}
    - \dfrac{d(d-2)}{4\ell^2} + m^2 \right] \right) \Phi = 0
\]
where $\partial_* = -2 \ell^{-1}\sqrt{u}f(u)\partial_u$
and $k_T^2=l(l+d-3)-2$ is the eigenvalue of the tensor harmonics
${\mathbb T}_{ij}$ on the $(d-2)$-sphere.
This is exactly the same equation as that for the massive Klein-Gordon field with mass-squared $m^2$ and the eigenvalue of the spherical harmonics $k^2=l(l+d-3)$. One might therefore anticipate that the stability criterion for generic tensor-type perturbations should coincide with that of the massive scalar field, $m^2<0$, which can be read off from, for example, the quasi-normal mode analysis of de Sitter spacetime~\cite{Lopez-Ortega:2006aal}. 
However, although the equations of motion for the two fields take the same form, a massive scalar field has a physical degree of freedom even for the S-wave ($l=0$ mode), while the tensor-type perturbations with $l=0$ become spurious modes, making it less straightforward to derive physical consequences concerning the stability. 
}. 

By the traceless and transverse conditions~(\ref{tt_cond}), we obtain the following three constraint equations 
\begin{subequations}
\label{constr_static}
\begin{align}
\label{constr1_static}
  & (d - 2)\, R + T + U = 0, 
\\
\label{constr2_static}
  & u\, S' - \frac{d+1}{2}\, S -i\, \Hat{\omega}\, u\, T
  =0, 
\\
\label{constr3_static}
  & u\, U' - \frac{1}{2}\, \left( d - 1 - \frac{1}{f} \right)\, U
  + \frac{ i\, \Hat{\omega} }{4\, f^2}\, S
  - \frac{1 + f}{2\, f}\, T
  = 0,
\end{align}
\end{subequations}
where $\hat{\omega}=\ell\omega$.
Eliminating $R$ from Eqs.~(\ref{H_equation}) by Eq.~(\ref{constr1_static}), another constraint equation 
is derived, 
\begin{align}
\label{constr4_static}
  & \left\{ \Hat{m}^2 + (d - 2)\, \Hat{\omega}^2 \right\}\, S
  + 2\, (i\, \Hat{\omega})\, \left\{ (\Hat{\omega}^2 + 1)\, u\, (T + U)
    - (\Hat{m}^2 - d + 2)\, f\, U
  \right\}
  =0. 
\end{align}
Eliminating $T$ and $R$ from Eqs.~(\ref{constr_static}) and (\ref{constr4_static}), two coupled first order 
differential equations for $(S, U)$ are derived. Introducing new variable $Z=S+2i\hat{\omega}(1-u)U$, the two equations for $(S,U)$ are reduced to 
the following single second order differential equation 
\begin{align}
\label{Z_dS_master}
  & Z'' + \left( \frac{1}{f} - \frac{d+1}{2\, u} \right)\, Z'
  + \frac{ \Hat{\omega}^2\, u - \left\{ \Hat{m}^2 + 2\, (d + 1) \right\}\, f }
         {4\, u^2\, f^2}\, Z
  = 0.
\end{align}

Setting $\Phi := u^{-(d+2)/4} Z$, we rewrite the above equation (\ref{Z_dS_master}) as
\begin{subequations}
\begin{align}
  & \hat{\omega}^2 \Phi = A \Phi := - 2\sqrt{u}f\dfrac{d}{du}\left( 2\sqrt{u}f\dfrac{d\Phi}{du} \right) + V(u) \Phi \,, \quad 
\\
  & V(u) := \dfrac{(d+2)(d+4)}{4}\dfrac{f^2}{u}- (d+2)ff' + \dfrac{f}{u}\left[ 2(d+1) + \hat{m}^2 \right] \,.
\end{align}
\end{subequations} 
Under the inner product defined by 
\bena
 (\Phi_1, \Phi_2):=\int \dfrac{du}{2\sqrt{u}f}\Phi_1^*(u)\Phi_2(u) \,, 
\eena
the operator $A$ is formally self-adjoint. Let us introduce the derivative operator $D$ by 
$D \Phi=2\sqrt{u}fG\partial_u (\Phi/G)$ with some function $G(u)$. Then, for any $\Phi \in C^\infty_0(1<u<\infty)$, i.e., $\Phi$ is a smooth function of compact support in the interval $(1,\infty)$, we find 
\bena
 \hat{\omega}^2\| \Phi \|^2= (\Phi, A\Phi)= \|D\Phi\|^2 + \int_1^\infty \dfrac{du}{2\sqrt{u}f} \tilde{V} |\Phi|^2 \,, 
\eena
where
\bena
 \tilde{V}:= V- 2\sqrt{u}f G^{-1}\partial_u(2\sqrt{u}f\partial_uG) \,.
\eena
It follows that if $\tilde{V}$ becomes positive definite for some $G$, then $A$ must also be positive definite, implying $\hat{\omega}^2>0$. If this is the case, $\omega$ is real and there is no exponentially growing unstable mode.  

Now we choose $G=u^{-(d+2)/4}$. Then we have
\bena  
 \tilde{V} =f (u)\left\{ \hat{m}^2+2(d+1)\right\} \,.
\label{stability-criterion}
\eena 
This is positive for $\hat{m}^2 \geqslant  -2(d+1)$. 

This leaves open the possibility of the existence of unstable modes for $\hat{m}^2<-2(d+1)$. We will show that precisely for this case $\hat{m}^2<-2(d+1)$, there exists an exponentially growing unstable mode by explicitly constructing unstable mode below. We should also note that even for $\hat{m}^2 \geqslant -2(d+1)$, the argument above does not eliminate the possibility of linear growth of the perturbations with respect to the static Killing time $t$.

\subsubsection{Unstable modes in static chart} 
Now we explicitly construct an unstable mode solution. For that purpose, we impose the outgoing wave boundary condition at the horizon, $u=1$. The solution satisfying the boundary condition is expressed by the hypergeometric function as  
\begin{subequations}
\begin{align}
\label{Z_sol1}
  & Z = (u-1)^{ -\frac{i}{2}\hat{\omega}} u^{(d+3-2p)/4} 
F(\alpha,\,\beta, \,1-i\hat{\omega};1-u), 
\\ 
  & \alpha := -\frac{1}{2}\left(p-\frac{d+3}{2}+i\hat{\omega} \right),
  \hspace{2.0truecm}
\beta:=-\frac{1}{2}\left(p+\frac{d-1}{2}+i\hat{\omega}  \right). 
\end{align}
\end{subequations} 
From the transformation of hypergeometric function, we have  
\bena
\label{Gauss_trans}
 F(\alpha,\beta,\gamma;1-u)
 &=&
\frac{\Gamma(\gamma)\Gamma(\beta-\alpha)}{\Gamma(\beta)\Gamma(\gamma-\alpha)}
u^{-\alpha}F\left(\alpha,\gamma-\beta,\alpha-\beta+1;\frac{1}{u}\right) 
 \nonumber \\
&{}& +\frac{\Gamma(\gamma)\Gamma(\alpha-\beta)}{\Gamma(\alpha)\Gamma(\gamma-\beta)}
u^{-\beta}F\left(\beta,\gamma-\alpha,\beta-\alpha+1;\frac{1}{u}\right) , 
\eena 
and we find that $Z$ behaves near the origin, $u=\infty$ as 
\begin{align}
Z\simeq C_1+C_2u^\frac{d+1}{2}. 
\end{align}
By the formula (\ref{Gauss_trans}), the regularity condition at $u=\infty$ leads to, 
\begin{align}
\label{regularity}
-n= 
\left\{
\begin{array}{lll}
\gamma-\beta & \longrightarrow & \hat{\omega}= i\left(2n+\dfrac{d+3}{2}+p   \right) \\ 
\alpha & \longrightarrow & \stackrel{\mathstrut}{\hat{\omega}}
  =-i\left(2n+\dfrac{d+3}{2}-p   \right) \\
\end{array}
\right.,  
\end{align}
where $n$ is a non-negative integer. From the latter condition combined
with (\ref{def-p}),
it follows that 
the imaginary part of $\hat{\omega}$ becomes positive for $n=0$ when
\begin{align}
\label{instability_dS_static}
\hat{m}^2<-2(d+1). 
\end{align} 
This implies that de Sitter spacetime is unstable in the static chart when the mass is negatively large. 
However, we should note that the static chart cannot cover the whole de Sitter spacetime, (i.e., there always exists a horizon for any static patch associated with a timelike geodesic observer), hence the above parameter range (\ref{instability_dS_static}) for the existence of unstable modes apply only to the analysis performed in the static patch. For this reason, we will examine the semiclassical (in)stability of de Sitter spacetime in the different chart below.   

\subsubsection{Stability analysis in the cosmological charts}
We next consider cosmological perturbations in the flat, closed, and open charts, respectively. 
Since the background spacetime is time-dependent, it is generally difficult to demonstrate the instability of perturbations directly. Therefore, in what follows, we examine whether the perturbation amplitudes grow 
relative to the background metric $\bar{\gamma}_{ij}$.

We express our background de Sitter spacetime in cosmological chart with the conformal time $\eta$ as follows: 
\bena
\label{metric_flat_chart}
 ds_d^2 &=& 
\overline{\bdyg}_{\mu\nu} dx^\mu dx^\nu 
= \ell^2a^2(\eta) \left(-d\eta^2+\overline{\gamma}_{\ij}dx^i dx^j\right) \,, 
\quad x^\mu=(\eta,\, x^i), 
\eena
where $\overline{\gamma}_{ij}$ denotes the $(d-1)$-dimensional metric of a maximally symmetric space with constant sectional curvature $K=0, \pm 1$ and $a(\eta)$ is the scale factor given by 
\begin{align}
 a(\eta) = \left\{
\begin{array}{ll}
-\dfrac{1}{\eta} & (K=0, \quad -\infty<\eta<0) \\
-\dfrac{1}{\sinh \eta} & (K=-1, \quad -\infty<\eta<0) \\
-\dfrac{1}{\sin\eta} & (K=1, \quad -\pi<\eta<0) 
\end{array}
\right.  \,.
\end{align}
Here, $\eta=-\infty$ corresponds to the null hypersurface for $K=0$ and $K=-1$. 

Let us consider the perturbed metric $ \overline{\bdyg}_{\mu\nu} +\epsilon H_{\mu\nu}$. The perturbed metric $H_{\mu\nu}$ can be decomposed as
\begin{subequations}
\begin{align}
&{H_0}^0={q_0}^0,
\\
& H_{0i}={\bm \nabla}_i q_0+q^{(1)}_{0i},
\\
& H_{ij}=q_L\overline{\gamma}_{ij}+P_{ij}\, q_T^{(0)}+2{\bm \nabla}_{(i}q_{Tj)}^{(1)}+q_{Tij}^{(2)},
\\
& P_{ij}:={\bm \nabla}_{(i}{\bm \nabla}_{j)}-\frac{\overline{\gamma}_{ij}}{d-1}{\bm \nabla}^2, 
\end{align}
\end{subequations} 
where ${\bm \nabla}_i$ denotes the covariant derivative with respect to $\overline{\gamma}_{ij}$, and
$q^{(1)}_{0i}$, $q_{Ti}^{(1)}$, and $q_{Tij}^{(2)}$ satisfy
\begin{align}
  & {\bm \nabla}^iq^{(1)}_{0i}={\bm \nabla}^iq_{Ti}^{(1)}=0,
& & {\bm \nabla}^jq_{Tij}^{(2)}=q^{(2)i}_{Ti}=0,  
\end{align} 
where the indices are lowered and raised by $\overline{\gamma}_{ij}$ whenever both indices refer to spatial coordinates.

For the tensor perturbation $q_{Tij}^{(2)}$, Eqs.~(\ref{H_equation}) reduce to 
\begin{align}
\label{tensor_cos}
\left\{a^{-d}\p_\eta\left(a^{d-2}\p_\eta  \right)+a^{-2}(-{\bm \nabla}^2+2K)+m^2\right\}
(a^{-2}q_{Tij}^{(2)})=0. 
\end{align}
By separating variables as $q_{Tij}^{(2)}= a^{2}(\eta) f_T(\eta){\bm T}_{ij}({\bm x})$, the spatial function 
${\bm T}_{ij}({\bm x})$ satisfies
\begin{align}
({\bm \nabla}^2+k_T^2){\bm T}_{ij}({\bm x})=0, 
\end{align}
where $k_T^2$ takes continuous non-negative values for $K=-1,0$, and discrete values
\begin{align}
k_T^2=l(l+d-2)-2, \qquad l=0,1,2,\cdots 
\end{align}
for $K=1$. Thus, Eq.~(\ref{tensor_cos}) becomes
\begin{align}
\label{Eq_f_T}
\ddot{f_T}+(d-2)\dfrac{\dot{a}}{a} \dot{f_T}+\left(k_T^2+2K+a^{2}\hat{m}^2\right)f_T=0, 
\end{align}
where the dot denotes differentiation with respect to $\eta$. 
The solutions can be expressed in terms of the Bessel function of the first kind $J_\nu(z)$ and the 
hypergeometric function $F(\alpha,\beta,\gamma;z)$ as follows:
for $K = 0$,
\begin{subequations}
\label{f_T_sol}
\begin{align}
\label{f_T_sol_K=0}
& f_T(\eta)=\eta^\frac{(d-1)}{2}\left[c_1J_{p}(k_T\eta)+c_2J_{-{p}}(k_T\eta)   \right],
\\
& p=\sqrt{\frac{(d-1)^2}{4}-\hat{m}^2},
\end{align} 
\end{subequations}
and for $K=\pm 1$, we obtain 
\begin{subequations}
\begin{align}
\label{f_T_sol_K=pm}
  & f_T(\eta)
  = c_1a^{p- {(d-1)}/{2}}F\left(\frac{1/2-\kappa-p}{2}, \frac{1/2+\kappa-p}{2},1-p;\frac{K}{a^2} \right)
\nonumber \\
  &\hspace*{1.0truecm}
  + c_2a^{-p-{(d-1)}/{2}}F\left(\frac{1/2-\kappa+p}{2},
    \frac{1/2+\kappa+p}{2},1+p;\frac{K}{a^2} \right), \\ 
& \kappa:=\sqrt{\frac{k_T^2}{K}+3-d+\frac{d^2}{4}}.
\end{align} 
\end{subequations}
When the mass-squared is negative $\hat{m}^2<0$, in the asymptotic region $\eta\to 0$, $f_T$ diverges as $\sim \eta^{(d-1)/2-p} \sim a(\eta)^{p-(d-1)/2}$, which is faster than the background de Sitter expansion $a(\eta)$, since $(d-1)/2<p$ for $\hat{m}^2 <0$. This behavior is independent of $K$, provided that $c_1\neq 0$. Since $f_T\sim {\delta \bdyg_i}^j$, the de Sitter spacetime is therefore unstable against tensor-type perturbations.

For the $K=1$ case, using the formula
\bena
 F(\alpha,\beta,\gamma;z)
 &=& \frac{\Gamma(\alpha+\beta-\gamma)\Gamma(\gamma)}
{\Gamma(\alpha)\Gamma(\beta)}(1-z)^{\gamma-\alpha-\beta}
F(\gamma-\alpha,\gamma-\beta,\gamma-\alpha-\beta+1;1-z) \nonumber \\
&{}&+\frac{\Gamma(\gamma-\alpha-\beta)\Gamma(\gamma)}
{\Gamma(\gamma-\alpha)\Gamma(\gamma-\beta)}
F(\alpha, \beta, \alpha+\beta-\gamma+1;1-z), 
\eena 
the perturbation remains regular at $\eta=-\pi/2$, since $\gamma-\alpha-\beta=1/2$.
This implies that spatially compact, initially regular data in the global coordinates lead to instability.

For the vector perturbations, ($q^{(1)}_{0i}$, $q^{(1)}_{Ti}$), the transverse conditions~(\ref{tt_cond}) 
lead to 
\begin{align}
\label{Constr_vec}
a^{-d}\p_\eta(a^{d}q_i^{(1)0})+a^{-2}\{\bm\nabla^2+(d-2)K   \}q_{Ti}^{(1)}=0. 
\end{align}  
From Eqs.~(\ref{H_equation}) for $(\mu, \nu) = (0,i)$ and $(i,j)$, together with Eq.~(\ref{Constr_vec}), we 
obtain two coupled equations,
\begin{subequations}
\label{vec_Eqs}
\begin{align}
& -a^{-2} \{{\bm\nabla}^2+(d-2)K  \}V_i+m^2q_i^{(1)0}=0,
\\
& a^{-d}\p_\eta(a^{d-2}V_i)+m^2(a^{-2}q_{Ti}^{(1)})=0, 
\end{align}
\end{subequations} 
where 
\begin{align}
\label{def_gauge_vec}
V_i:=q_i^{(1)0}+\p_\eta(a^{-2}q_{Ti}^{(1)}). 
\end{align}
By using Eq.(\ref{vec_Eqs}),
both $q^{(1)}_{0i}$ and $q^{(1)}_{Ti}$ can be expressed 
in terms of $V_i$. Eliminating these variables from Eq.(\ref{def_gauge_vec}),
we obtain the master equation for the 
new variable $V_i$ as
\begin{align}
\label{vec_master}
\left[a^{d-2}\p_\eta(a^{-d}\p_\eta)-a^{-2}\{{\bm\nabla}^2+(d-2)K\}+m^2\right]
(a^{d-2}V_i) = 0. 
\end{align}

In the asymptotic region $\eta \to 0$, the scale factor behaves as $a \sim 1/\eta$.
Then Eq.~(\ref{vec_master}) leads to the asymptotic solution
\begin{align}
V_i\sim \eta^{\frac{(d-1)}{2}-1-p}. 
\end{align}
The asymptotic behavior of $q^{(1)}_{0i}$ and $a^{-2}q^{(1)}_{Ti}$ is given by
\begin{align}
q_i^{(1)0}\sim \eta^{\frac{d-1}{2}+1-p}, \qquad a^{-2}q^{(1)}_{Ti}\sim \eta^{\frac{d-1}{2}-p}. 
\end{align}
This shows that the latter variable becomes more unstable and asymptotically grows when $\hat{m}^2 < 0$.

For the scalar perturbations, the traceless and transverse conditions~(\ref{tt_cond}) reduce to 
\begin{subequations} 
\label{Constr_scalar}
\begin{align}
\label{Constr_scalar1}
  & {q_0}^0+(d-1)a^{-2}q_L=0, 
\\
\label{Constr_scalar2}
  & (d-1) a^{-(d+1)} \p_\eta (a^{d-2}q_L) + a^{-2}{\bm\nabla}^2(aq^0)=0, 
\\
\label{Constr_scalar3}
  & a^{-d} \p_\eta (a^{d}q^0) + a^{-2} q_L + \frac{d-2}{d-1} ({\bm \nabla}^2+(d-1)K) (a^{-2}q_T^{(0)})=0. 
\end{align}
\end{subequations}
Substituting $\mu = \nu = 0$ into Eqs.~(\ref{H_equation}), we also obtain 
\begin{align}
    \left[
       a^{-(d+2)} \p_\eta (a^{d}\p_\eta) -a^{-2} {\bm \nabla}^2 + m^2 + \dfrac{2}{\ell^2} 
   \right]{q_0}^0
  +2 \dot{a}a^{-3}{\bm\nabla}^2q^0=0. 
\end{align}
Eliminating $q^0$ using Eqs.~(\ref{Constr_scalar1}) and (\ref{Constr_scalar2}), we finally obtain
\begin{align}
\label{master_scalar}
 \left[
      a^{-(d+4)} \p_\eta (a^{d+2}) \p_\eta 
  - a^{-2} \{ {\bm\nabla}^2+2dK \}
      +m^2+\frac{2(d+1)}{\ell^2}    
 \right]{q_0}^0=0. 
\end{align}

As in the tensor and vector cases, the quantities $a^{-2}q_L$ and $a^{-2}q_T^{(0)}$ should be compared with the background metric $\bar{\gamma}_{ij}$. Using Eqs.~(\ref{Constr_scalar}) and (\ref{master_scalar}), the asymptotic behavior near $\eta = 0$ is found to be
\begin{align}
a^{-2}q_L\sim {q_0}^0\sim \eta^{\frac{d-1}{2}+2-p}, \qquad a^{-2}q_T^{(0)}\sim  \eta^{\frac{d-1}{2}-p}. 
\end{align}
This indicates that the latter variable is more unstable and asymptotically grows when $\hat{m}^2 < 0$.

We have shown that the cosmological perturbations grow indefinitely toward $\eta = 0$ whenever $\hat{m}^2 < 0$. 
Our unstable mode shows a power-law behavior in the conformal time $\eta$, which is transformed to exponential grow in the cosmic proper time. 
It is noteworthy that the Kretschmann scalar, $\bdyR_{\mu\alpha\beta\rho}\bdyR^{\mu\alpha\beta\rho}$, 
diverges on the null hypersurfaces at $\eta = -\infty$ for the $K = 0$ and $K = -1$ cases.
On the other hand, for $K = 1$, the coordinate chart covers the entire de Sitter spacetime, and the perturbation 
remains regular at the Cauchy surface $\eta = -\pi/2$.
Since the solutions are time-symmetric with respect to $\eta = -\pi/2$, the initial Cauchy data with small 
perturbations at $\eta = -\pi/2$ lead to an instability in both the future and past directions.

We have previously obtained the unstable mode of exponential growth in the static chart time $t$ when $\hat{m}^2< -2(d+1)$. We suspect that the mismatch of the parameter ranges for the instabilities between the cosmological and the static charts stems from the fact that the static chart is limited by the de Sitter (cosmological) horizon and never covers the outside the horizon\footnote{
Note that the stability criterion in the static chart relies on the restricted class of metric perturbations, which thus cannot form a complete set of basis vectors for generic perturbations in the static chart. 
Even if a complete orthonormal basis in the static chart is available at our hand, we still would not be able to express general perturbations in the cosmological chart, as the latter covers a larger portion of the de Sitter spacetime than the static chart. In addition, the unstable modes in the cosmological chart are homogeneous perturbations on the cosmological slice (e.g., $a^{-2}q_L$ and $a^{-2}q_T^{(0)}$ above), and hence fail to be normalizable with respect to the inner product on the static slice. This discrepancy may also account for the mismatch (This situation is a reminiscent of the Rindler modes in Minkowski spacetime.).
}.
On the other hand, the power-law instability found in the cosmological chart appears asymptotic region near the future infinity, occurring always outside of the cosmological horizon $\dot{J}^-(\lambda)$ for any complete timelike geodesic observer $\lambda$. Therefore, the stability criterion for the static chart given below (\ref{stability-criterion}) does not apply to the cosmological case.

\subsection{Minkowski spacetime}
In this subsection, we show that semiclassical perturbations obeying Eqs.~(\ref{H_equation}) with negative mass $m^2<0$ are unstable in Minkowski spacetime of any dimension. Since the background curvature vanishes, Eqs.~(\ref{H_equation}) reduce to the massive scalar field equation,
\begin{align}
\label{massive_scalar}
(\partial_x^2 - m^2)\phi(x) = 0,
\qquad x^\mu = (t,\bm{x}).
\end{align}
We introduce the invariant delta function $\Delta_d(x)$. The invariant delta function satisfies
$\Delta_d(x) = - \Delta_d(-x)$ and
\begin{subequations}
\label{def:delta_func}
\begin{align}
  & (\p_x^2 - m^2)\, \Delta_d(x) = 0,
\\
  & \Delta_d(x)\, \big|_{t=0} = 0,
& & \p_t \Delta_d(x)\, \big|_{t=0} = - \delta^{(d-1)}(\bm{x}),
\end{align}
\end{subequations} 
where $\delta^{(d-1)}(\bm{x})$ is the $(d-1)$-dimensional delta function. Then, the solution of Eq.~(\ref{massive_scalar}) with arbitrary regular initial data 
$(\varphi(x_0), \p_{t_0} \varphi(x_0))$~($x^\mu_0 = ( t_0,\, \bm{x}_0 )$) can be written as
\begin{align}
\label{sol_scalar_wave}
  & \phi(x)
  = \int d^3x_0~\Big[\, \big\{ \p_{t_0}\Delta_d(x - x_0) \big\}\, \varphi(x_0)
  - \Delta_d(x - x_0)\, \p_{t_0}\varphi(x_0)\, \Big]
  ~. 
\end{align}
This representation shows that the stability properties of the perturbation are determined entirely by the invariant delta function $\Delta_d(x)$. 

Since we are interested in the negative mass-squared case $m^2<0$, we present the delta functions 
$\Delta_d(x)$ for $d=3,4,5$ (see Appendix~\ref{Delta_function} for details):
\begin{subequations}
\begin{align}
  & \Delta_3(x)
  = - \frac{\text{sgn}(t)}{2\, \pi}\, \theta(\sigma^2)\,
  \frac{\cosh(|m|\, \sigma)}{\sigma}, 
\\
  & \Delta_4(x)
  = - \frac{\text{sgn}(t)}{2\pi}\,
  \left(\delta(\sigma^2) + \frac{|m^2|}{2}\, \theta(\sigma^2)\,
  \frac{I_1(|m|\, \sigma)}{|m|\, \sigma} \right),
\\
  & \Delta_5(x)
  = - \frac{\text{sgn}(t)}{ 2\, \pi^2\, \sigma }
  \left[\, \delta(\sigma^2)\, \cosh(|m|\, \sigma)
  + \frac{|m|}{ 2\, \sigma }\, \theta(\sigma^2)\,
  \left\{ \sinh(|m|\, \sigma)
    - \frac{\cosh(|m|\, \sigma)}{|m|\, \sigma}
  \right\}\, \right]
  , 
\end{align}
\end{subequations}
where $\sigma:=\sqrt{t^2 - r^2}$~($r := |\, \bm{x}\, |$). 

Now consider the solution~(\ref{sol_scalar_wave}) with regular compactly supported initial data $(\varphi(t_0, {\bm x}), \p_t\varphi(t_0, {\bm x}))$. For sufficiently large $\sigma$, i.e., when $t\gg r$, all of the above solutions 
diverge exponentially as $\Delta_d\sim e^{|m|\sqrt{\sigma^2}}$ for $d=3,4,5$. Therefore, Minkowski spacetime is always 
unstable whenever the system of Eqs.~(\ref{xi_equation}) admits a mode with negative mass squared $m^2<0$.

\section{Summary and discussions}\label{sec:6}
We have investigated the stability of $d$-dimensional ($d=3,4,5$) de Sitter and Minkowski spacetimes within the framework of semiclassical gravity, where the source term is given by a strongly coupled quantum field with a gravity dual.
The perturbed bulk Einstein equations are decomposed into the Lichnerowicz equations~(\ref{H_equation}) with mass-squared term and the bulk radial equation~(\ref{xi_equation}), where the semiclassical Einstein~(SCE) equations are encoded in a dynamical boundary condition at the AdS boundary. 
As shown in section~\ref{sec:5}, the negative mass-squared solutions of Eq.~(\ref{H_equation}) always lead to instabilities of the background de Sitter and Minkowski spacetimes.
As summarized in Table~\ref{table:1}, we have found that our stability results depend on the dimensions $d$. 
In particular, as we have shown in (\ref{cond_gammma_5_dS}) and (\ref{cond_gammma_5_Min}), the existence of negative mass-squared solutions for $d=5$ imposes a strong restriction on the allowed range of the dimensionless 
parameter $\gamma_5$, typically requiring $\gamma_5\lessapprox \beta^{3/2}$. However, 
in such negative-mass-squared solutions, $|\, m^2\, | \sim 1/|\, \alpha_i\, |$, indicating that the magnitude of the 
higher-curvature terms is of the same order as that of the Einstein tensor, and hence the perturbative analysis is unreliable. 

Our results are restricted to maximally symmetric spacetimes. It would be interesting to investigate whether they can be generalized to homogeneous but anisotropic cosmological models such as the Taub-NUT spacetime, or to black hole spacetimes, including higher-dimensional cases such as black strings.
In contrast to the AdS holographic semiclassical black hole case~\cite{Hamaki:2025mde}, one must impose an additional boundary condition at the cosmological horizon in asymptotically de Sitter spacetimes, or at infinity in asymptotically flat semiclassical black holes. For a specific value of the background black hole parameter, such as the mass, both boundary conditions---at the event horizon and at the cosmological horizon (or at infinity)---could be simultaneously satisfied. This tuning corresponds to the 
existence of regular semiclassical modes, giving rise to a possible hairy black hole solution.
It would then be worth investigating the thermodynamical stability of such hairy black holes.

Another interesting direction is to explore whether the dynamical boundary condition in our semiclassical approach can be realized in the brane-world holographic model, where the bulk gravity interacts with a higher-derivative theory of gravity coupled to a cut-off CFT on the brane.
Although one might naively expect that pushing the brane toward the AdS boundary would realize such a dynamical boundary condition, it has been reported that this setup may not provide a smooth interpolation between the brane-world holography and the dynamical-boundary frameworks~\cite{Llorens:2025sxw}. We hope that these investigations will provide deeper insights 
into the nature of semiclassical spacetimes and the interplay between quantum fields and geometry.

\begin{center}
{\bf Acknowledgments}
\end{center} 
We wish to thank Bob Wald for valuable discussions on the stability analysis of de Sitter space. 
This work is supported in part by JSPS KAKENHI Grant No. 25K07306 (K.M.) and also supported 
by MEXT KAKENHI Grant-in-Aid for Transformative Research Areas A Extreme Universe 
No.21H05186 (A.I. and K.M.) and 21H05182.

\appendix
\section{Explicit form of $\tau^{(d)}_{\mu\nu}$}\label{tau}
The boundary tensor $\tau^{(d)}_{\mu\nu}$ is derived from the variation of $\Gamma^{(d)}$~(\ref{bulkct}) as
\begin{subequations}
\label{vev_SE_tau}
\begin{align}
\label{vev_SE_tau_d=3}
\tau^{(3)}_{\mu\nu}&=0,
\\
\label{vev_SE_tau_d=4}
\tau^{(4)}_{\mu\nu}&=-\frac{L^3}{64\pi G_5}\,\Biggl\{\left(\tilde{D}^2-\frac{2}{3}\tilde{R}\right)
\tilde{R}_{\mu\nu}-\frac{1}{3}\tilde{D}_\mu \tilde{D}_\nu \tilde{R}+2\tilde{R}_{\alpha\mu\beta\nu}
\tilde{R}^{\alpha\beta} \nonumber \\
&\hspace*{2.0truecm}
-\frac{\tilde{g}_{\mu\nu}}{2}
\left(\tilde{R}_{\alpha\beta}\tilde{R}^{\alpha\beta}+\frac{1}{3}\tilde{D}^2\tilde{R}
-\frac{1}{3}\tilde{R}^2\right)  \Biggr\}\times \ln \left(\mu^2z^2\right), 
\\
\label{vev_SE_tau_d=5}
\tau^{(5)}_{\mu\nu}&=\frac{L^3}{72\pi G_6\Omega}\Biggl\{\left(\tilde{D}^2-\frac{5}{8}\tilde{R}\right)\tilde{R}_{\mu\nu}
-\frac{3}{8}\tilde{D}_\mu \tilde{D}_\nu \tilde{R}+2\tilde{R}_{\alpha\mu\beta\nu}\tilde{R}^{\alpha\beta}
\nonumber \\
&\hspace*{2.0truecm}
-\frac{\tilde{g}_{\mu\nu}}{2}
\left(\tilde{R}_{\alpha\beta}\tilde{R}^{\alpha\beta}+\frac{1}{4}\tilde{D}^2\tilde{R}
-\frac{5}{16}\tilde{R}^2\right)  \Biggr\}. 
\end{align}
\end{subequations}

\section{Perturbations of Maximally symmetric spacetime} 
In $d$-dimensional dS~($k=1$), Minkowski~($k=0$), and AdS~($k=-1$) spacetime backgrounds, one obtains the first 
order perturbations of the curvature tensor and the derivatives under the transverse-traceless 
conditions~(\ref{tt_cond}) and Eqs.~(\ref{H_equation}). The result is 
\begin{subequations}
\label{formula_dS}
\begin{align}
& \delta \bdyR^{\mu\alpha}_{\lambda\sigma}=-\epsilon\left(2\overline{\bdyD}_{[\lambda}
\overline{\bdyD}^{[\mu}{H^{\alpha]}}_{\sigma]}
 +k\frac{2}{\ell^2}{\delta^{[\mu}}_{[\lambda}{H^{\alpha]}}_{\sigma]}\right),
& & \delta {\bdyR^\alpha}_{\sigma}=-\epsilon\frac{m^2}{2} {H^\alpha}_\sigma,
\\
& \delta ({\cal \overline{D}}_\rho{\bdyR^\alpha}_{\sigma})=-\epsilon\frac{m^2}{2} {\cal \overline{D}}_\rho{H^\alpha}_\sigma, 
\\
& \delta ({\cal \overline{D}}^2{\bdyR^\alpha}_{\sigma})=-\epsilon\frac{m^2}{2} {\cal \overline{D}}^2{H^\alpha}_\sigma, 
& & \delta ({\cal \overline{D}}_\mu{\cal \overline{D}}^\nu \bdyR)=0. 
\end{align}
\end{subequations} 
The perturbations of the higher curvature corrections $\mathcal{H}^{(i)}_{\mu\nu}$ in Eqs.~(\ref{def:H}) and 
the tensor $\tau^{(i)}_{\mu\nu}$~($i=4,5$) in Eqs.~(\ref{vev_SE_tau})
are derived from Eq.(\ref{formula_dS}) as
\begin{subequations}
\label{H_pert_dS}
\begin{align}
\label{H1_pert_dS}
  & \delta {\cal H}_\mu^{(1)\nu}=-k\frac{d(d-1)}{\ell^2}m^2{H_\mu}^\nu, 
\\
\label{H2_pert_dS}
  & \delta {\cal H}_\mu^{(2)\nu}=-\frac{m^2}{2}\left(m^2+\frac{2k(d-1)}{\ell^2} \right){H_\mu}^\nu, 
\\
\label{H3_pert_dS}
  & \delta {\cal H}_\mu^{(3)\nu}=-2m^2\left(m^2-\frac{k(d-4)}{\ell^2} \right){H_\mu}^\nu, 
\end{align}
\end{subequations}
and
\begin{subequations}
\label{pert_tau}
\begin{align}
\label{pert_tau4}
  & \delta \tau^{(4)}_{\mu\nu}=\frac{L^3}{64\pi G_5}\left(\frac{m^4}{2}-k\frac{m^2}{\ell^2}  \right) 
h_{\mu\nu}\times \ln \left(\mu^2z^2\right), 
\\
\label{pert_tau5}
  & \delta \tau^{(5)}_{\mu\nu}=\frac{L^3}{72\pi G_6\Omega}
\left(\frac{9}{2\ell^4}-\frac{m^4}{2}+\frac{9km^2}{4\ell^2}  \right)h_{\mu\nu}. 
\end{align}
\end{subequations}
Substituting Eqs.~(\ref{formula_dS}) and (\ref{H_pert_dS}) into Eqs.~(\ref{semi_Eqs}),  we obtain 
\begin{align}
\label{Pert_Ein_dS} 
\delta (\mathcal{E}_\mu{}^\nu)=-\frac{\epsilon\, \hat{m}^2}{2\ell^2}
\left[1+2k(d-1)(d\hat{\alpha}_1+\hat{\alpha}_2)+\hat{m}^2(\hat{\alpha}_2+4\hat{\alpha}_3)
-4k(d-4)\hat{\alpha}_3 \right]{H_\mu}^\nu, 
\end{align}
where we must set $\alpha_i=0$ for $d=3$. 

\section{The invariant delta functions $\Delta_d(x)$} 
\label{Delta_function}
For the positive mass-squared case $m^2>0$, 
the invariant delta function $\Delta_d(x)$, which satisfies the properties~(\ref{def:delta_func}), can be written as
\begin{subequations}
\label{sol:delta_func}
\begin{align}
 & \Delta_d(x)
  = - \int \frac{d^{d-1}q}{(2\pi)^{d-1}}~\frac{\sin(\omega_q t)}{\omega_q}\,
    e^{ i \bm{q}\cdot \bm{x} },
& & x^\mu=(t,\,\bm{x}), 
\label{sol:delta_func-I} \\
  &\hspace*{1.0truecm}
  = - i \int \frac{d^dk}{(2\pi)^{d-1}}~\text{sgn}(k^0)\,
    \delta(k^2 + m^2)\, e^{i k \cdot x}
  ~,
\label{sol:delta_func-II} \\
  & \omega_q := \sqrt{q^2 + m^2}.
\end{align}
\end{subequations}
Here, $\text{sgn}(u):=u/|u|$ ($u\neq 0$) with $\text{sgn}(0)=0$ denotes the sign function, and 
$k^\mu=(k^0, {\bm q})$ is the $d$-dimensional momentum vector.
When $m^2>0$, the invariant delta function $\Delta_d(x)$ becomes Lorentz invariant, as expected from (\ref{sol:delta_func-II}). 

The invariant delta functions $\Delta_d(x)$ for $d=2,3$ evaluate to
\begin{subequations}
\label{delta_2_3_positive}
\begin{align}
  & \Delta_2(x)
  = - \int^\infty_{-\infty} \frac{dq}{2\, \pi}\,
    \frac{\sin(\omega_q t)}{\omega_q}\, \cos(q x)
  = - \frac{\text{sgn}(t)}{2}\, \theta(\sigma^2)\, J_0(m \sigma),
\\
  & \Delta_3(x)
  = - \int^\infty_0 \frac{dq}{2\, \pi}\, q\,
    \frac{\sin(\omega_q t)}{\omega_q}\, J_0(qr)
  = - \frac{\text{sgn}(t)}{2\, \pi}\, \theta(\sigma^2)\,
  \frac{\cos(m \sigma)}{\sigma}, 
\end{align} 
\end{subequations}
where $\theta(z)$ is the Heaviside step function, $J_n(x)$ denotes the $n$th-order Bessel function of the first kind, 
$r=|\bm x|$, and $\sigma=\sqrt{t^2-r^2}$.
These expressions can be immediately obtained by substituting $t=0$ or $r=0$, using the Lorentz invariance 
of $\Delta_d(x)$.

Similarly, for the negative mass-squared case we obtain
\begin{subequations}
\begin{align}
\label{sol:delta_func-tachyonic}
  & \Delta_d(x)
  = - \int_{q > |m|} \frac{d^{d-1}q}{(2\pi)^{d-1}}~
    \frac{ \sin(\omega_q t) }{\omega_q}\, e^{ i \bm{q}\cdot \bm{x} }
  - \int_{q < |m|} \frac{d^{d-1}q}{(2\pi)^{d-1}}~
    \frac{ \sinh( |\, \omega_q\, |\, t) }{|\, \omega_q\, |}\,
    e^{ i \bm{q}\cdot \bm{x} }
\\
  & \omega_q
  := \sqrt{ q^2 - |\, m\, |^2 }.
\end{align}
\end{subequations}
Thus, although the Lorentz invariance of $\Delta_d(x)$ is not apparent for $m^2 < 0$,  
by careful inspection of (\ref{sol:delta_func-tachyonic}) for $d=2,3$, we can obtain the following Lorentz invariant results:
\begin{subequations}
\label{delta_2_3_negative}
\begin{align}
  & \Delta_2(x)
  = - \frac{\text{sgn}(t)}{2}\, \theta(\sigma^2)\, I_0(|m|\, \sigma),
\\
  & \Delta_3(x)
  = - \frac{\text{sgn}(t)}{2\, \pi}\, \theta(\sigma^2)\,
    \frac{\cosh(|m|\, \sigma)}{\sigma}, 
\end{align} 
\end{subequations}
where $I_n(z)$ denotes the modified $n$th-order Bessel function of the first kind. 
The results agree with those obtained by analytically continuing the positive mass-squared expressions~(\ref{delta_2_3_positive}).

The higher-dimensional invariant delta functions $\Delta_d(x)$ for $d=4,5$ can be derived from the lower-dimensional ones~(\ref{delta_2_3_positive}) and (\ref{delta_2_3_negative}) as
\begin{subequations}
\label{delta_4_5}
\begin{align}
  & \Delta_4(x)
  = - \frac{1}{\pi\, r}\, \int^\infty_0 \frac{dq}{2\, \pi}\, q
    \frac{\sin(\omega_q t)}{\omega_q}\, \sin(q r)
  = - \frac{1}{2\, \pi\, r}\, \frac{\p}{\p r}\, \Delta_2(x),
\\
  & \Delta_5(x)
  = - \frac{1}{2\, \pi\, r}\,
    \int^\infty_0 \frac{dq}{2\, \pi}\, q^2\,
    \frac{\sin(\omega_qt)}{\omega_q}\, J_1(qr)
  = - \frac{1}{2\, \pi\, r}\, \frac{\p}{\p r}\, \Delta_3(x),
\end{align}
\end{subequations}
where in the last equality we used the identity $J_0'(z)=-J_1(z)$.


\end{document}